\documentclass[twoside,english,12pt]{cern2}

\usepackage{amssymb}

\usepackage{graphicx}
\usepackage{babel}
\usepackage{multicol}
\usepackage[centerlast]{caption2}
\usepackage{setspace}

\begin{document}

\title{Electron Conditioning of Technical Aluminium Surfaces: Effect on the Secondary
Electron Yield}
\author{F. Le Pimpec, F. King, R.E. Kirby \\
SLAC, 2575 Sand Hill Road, Menlo Park, CA 94025 \\}
\maketitle

\begin{abstract}
The effect of electron conditioning on commercially aluminium
alloys 1100 and 6063 were investigated. Contrary to the assumption
that electron conditioning, if performed long enough, can reduce
and stabilize the SEY to low values ($\leq 1.3$, value of many
pure elements \cite{crc}), the SEY of aluminium did not go lower
than 1.8. In fact, it reincreases with continued electron exposure
dose.
\end{abstract}

\footnotetext{Work supported by the Department of Energy Contract
DE-AC02-76SF00515}

\doublespacing

\section{Introduction}

In the framework of the ILC electron cloud suppression, studies on
secondary electron emission (SEE) from technical surfaces are
ongoing.

In this brief paper we will present secondary electron yield
(SEY), $\delta$, results obtained from two technical surfaces :
aluminium alloys 1100 and 6063. We compare these results to other
data obtained elsewhere.

It is known, from the literature, that a metallic aluminium
surface has a $\delta_{max}$ below 1 \cite{crc}. However, its
technical surface is oxidized, and the $\delta_{max}$ can be well
above 2.5. This value might not be compatible with the running of
a charged positively particle beam, and it becomes necessary to
find a way to lower the yield of such surfaces. Coatings and
electron or ion conditioning are two ways of achieving this goal
\cite{lepimpec:LCC128} \cite{lepimpec:LCC146}.

\section{Experiment Description and Methodology}
\label{expsetupdescript}

\begin{figure}[tbph]
\begin{center}
\includegraphics[width=0.75\textwidth,clip=]{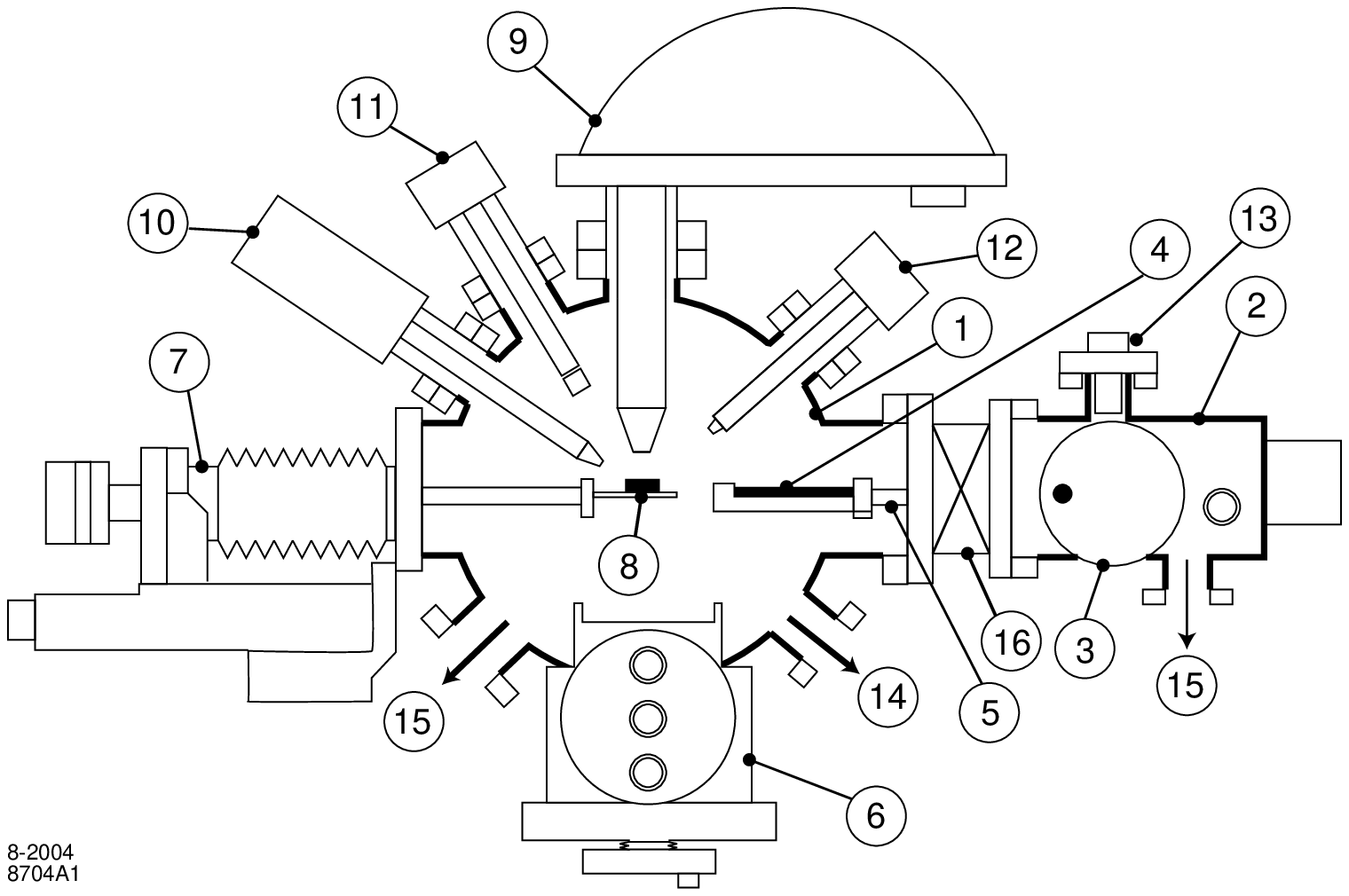}
\end{center}
\caption{Experimental setup} \label{figsketchsetup}
\end{figure}

The system, sketch in Fig.\ref{figsketchsetup}, and experimental
methodology used to measure the secondary electron yield have been
described thoroughly in \cite{lepimpec:LCC128}. Hence, we will
summarize the description of the SEY system.

\begin{enumerate}
\begin{minipage}[h]{.5\linewidth}
\item Analysis chamber
\item Loadlock chamber
\item Sample plate entry
\item Sample transfer plate
\item Rack and pinion travel
\item Sample plate stage
\item XYZ $\theta$ Omniax$^{TM}$ manipulator
\item Sample on XYZ $\theta$
\end{minipage}
\begin{minipage}[h]{.5\linewidth}
\item Electrostatic energy analyzer
\item X-ray source
\item SEY/SEM electron gun
\item Microfocus ion gun
\item Sputter ion gun
\item To pressure gauges and RGA
\item To vacuum pumps
\item Gate valve
\end{minipage}
\end{enumerate}

The system is composed of two coupled stainless steel ultra high
vacuum (UHV) chambers where the pressure is in the low 10$^{-10}$
Torr scale in the measurement chamber and high 10$^{-9}$ Torr
scale in the "load lock" chamber. Samples, individually screwed to
a carrier plate, are loaded first onto an aluminium transfer plate
in the load lock chamber, evacuated to the low 10$^{-8}$ Torr
scale, and then transferred into the measurement chamber.

The sample to be measured is installed on a special manipulator
arm. The feature of this arm allow us to bake the loaded sample,
and the temperature is recorded by the use of type C
thermocouples. The back of the samples are heated by electron
bombardment. This is achieved by biasing a tungsten filament
negatively.

\begin{figure}[tbph]
\begin{center}
\includegraphics[width=0.5\textwidth,clip=]{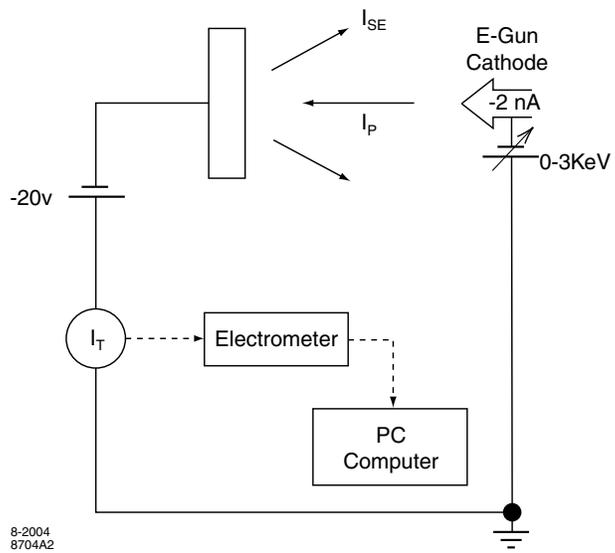}
\end{center}
\caption{Electronic circuitry used to measure the secondary
emission yield} \label{figelctrqcircuit}
\end{figure}

The electronic circuit for SEY measurement is that presented in
Fig.\ref{figelctrqcircuit} \cite{kirby:2001}. The energy of the
computer-controlled electron beam coming from the gun is decoupled
from the target measurement circuitry. However, the ground is
common to both. The target is attached to a bias voltage supply
and an electrometer connected in series to the data gathering
computer Analog Digital Converter (ADC). Measurements were made
with a Keithley 6487, a high resolution picoameter with internal
variable $\pm$505~V supply and IEEE-488 interface. The 6487 has
several filter modes which were turned off for our measurements.
The integration time for each current reading is 167~$\mu$s, which
is the minimum  value for the instrument. The current was sampled
one hundred times; the mean and standard deviation were returned
from the picoameter to the computer.

The SEY ($\delta$) definition is determined from
equation~\ref{equdefinition}. In practice equation~\ref{equSEY} is
used because it contains parameters measured directly in the
experiment.

\begin{minipage}[h]{.6\linewidth}
\begin{equation}
\delta = \frac{Number\ of\ electrons\ leaving\ the\
surface}{Number\ of\ incident\ electrons}
\label{equdefinition}
\end{equation}
\end{minipage}
\begin{minipage}[h]{.35\linewidth}
\begin{equation}
\delta = 1 -\frac{I_T}{I_P}
\label{equSEY}
\end{equation}
\end{minipage}

Where I$_P$ is the primary current (the current leaving the
electron gun and impinging on the surface of the sample) and I$_T$
is the total current measured on the sample ($I_T~=~I_P~+~I_S$).
I$_S$ is the secondary electron current leaving the target. The
reproducibility of the experiment is around 2\%.

\section{Effect of 130~eV Electron Bombardment on the SEY of Al}

\subsection{History of Aluminium 1100 and 6063 samples}

Aluminium 1100 is composed, at the minimum, of 99\% Al. Copper is
present in the range of 0.2\% to 0.5\%. The other elements,
present as impurities, are manganese, zinc, silicon and iron.
Aluminium 6063 is composed of 98.9\% Al, 0.45\% to 0.9\% of Mg,
and 0.2\% to 0.6\% of Si. Other impurities for Al 1100 are also
present in 6063, including copper.

The samples were chemically cleaned for UHV use, but not
deliberately passivated, and then kept in a dry nitrogen purged
box.

In an attempt to create an aluminium-nitride (AlN) thin film, for
the purpose of lowering the SEY, the Al 1100 sample was heated to
200$^\circ$C with pure hot 200$^\circ$C nitrogen gas blown on it.
The results were not encouraging (too low temperature), so the
sample surface was scraped clean in air with a tungsten carbide
tool, and was loaded in the SEY system. XPS confirmed that the
sample was quite clean, but air oxidized.

Two Al 6063 samples were also tested. The first sample will be
referred to as Al~6063. This sample was also scraped clean before
loading into the SEY system. The second sample, which is not
scraped, will be referred as the LER~Al~6063 sample. This sample
comes from a piece of the extruded Low Energy Ring (LER) chamber
made of Al 6063. This piece was UHV cleaned and stored in air for
years and then loaded as it in the SEY system. The LER accelerator
provides the positron beam in the SLAC B-factory.

\subsection{Secondary Electron Yield of Al 1100}

\begin{figure}[htbp]
\centering
\includegraphics[width=0.7\textwidth,clip=]{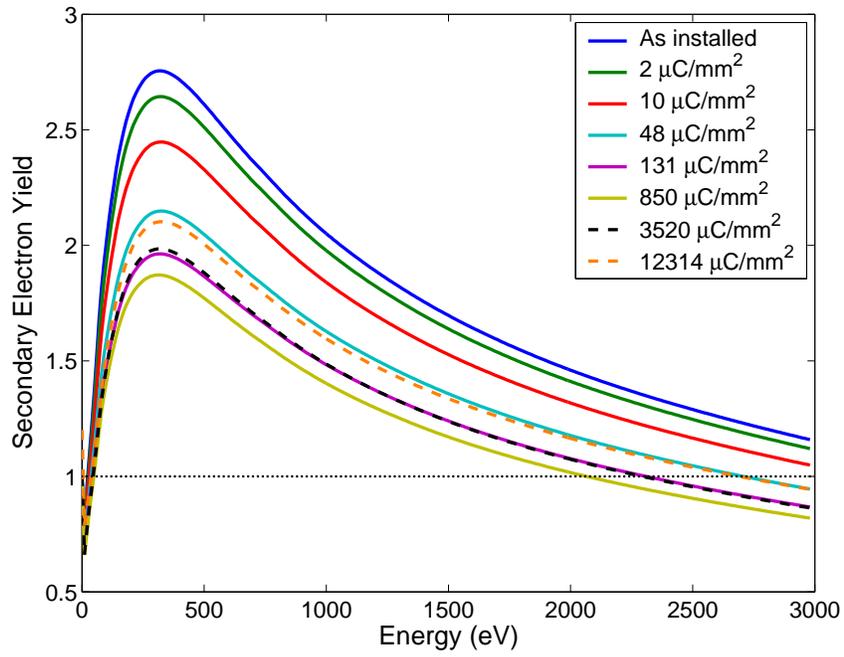}
\caption{Al 1100 exposed to electron conditioning. The primary
electron beam was impinging at 23$^\circ$ from normal incidence}
\label{figSEYal1100curves}
\end{figure}

The SEY results obtained by exposing the Al 1100 sample to an
electron conditioning beam of 130~eV kinetic energy are presented
in Fig.\ref{figSEYal1100curves} and \ref{figSEYvsdoseAl}. In the
NLC positron damping ring the average energy of the electrons from
the cloud was computed to be 130~eV \cite{lepimpec:LCC146}. The
SEY values were measured for a primary beam impinging the Al
surface at 23$^\circ$ from normal incidence. During the
conditioning, the pressure rose to 2.10$^{-9}$~Torr equivalent
N$_2$, due to electron stimulated desorption (ESD) from the
sample. As the dosing continued the pressure diminished to
5.10$^{-10}$~Torr. The effect of electron conditioning of ESD on
Al was, and still is, widely documented \cite{Ding:89}.

During the first 1000~$\mu$C/mm$^2$, the SEY of the Al 1100 sample
goes down as expected. However, we can see that this trend seems
to level off, suggesting that the conditioning of aluminium is a
very long process, Fig.\ref{figSEYvsdoseAl}. However, the next
point at 3520~$\mu$C/mm$^2$ shows an increase of the SEY, hence an
increase of the $\delta_{max}$.

\begin{figure}[htbp]
\centering
\includegraphics[width=0.6\textwidth,clip=]{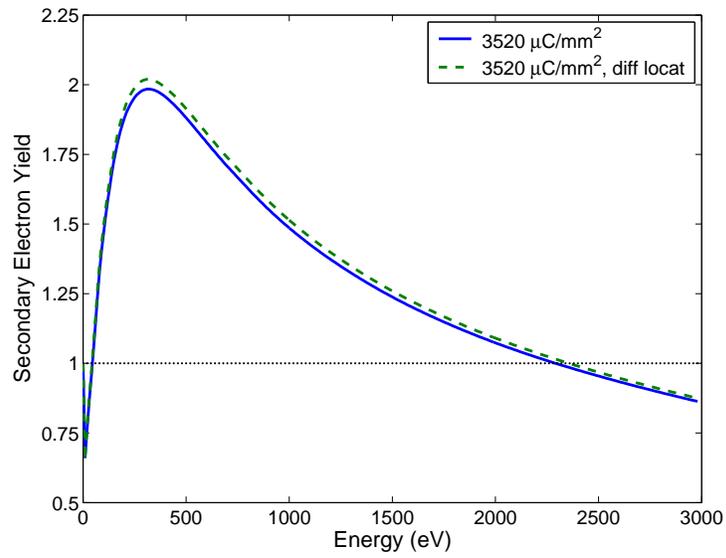}
\caption{SEY of Al 1100 at two different locations, same electron
dose of 3520~$\mu$C/mm$^2$} \label{figSEYtwolocations}
\end{figure}

This increase is contrary to expectation, i.e, normally yield
decreases with dose. In order to check the consistency of the
value, the sample was moved 5~mm, and a second point was
collected, Fig.\ref{figSEYtwolocations}. The results agreed and we
continued the conditioning to still higher dose.

To check the measurement system reproducibility, a previously
conditioned and measured NEG sample was installed on the holder
and re-measured. A NEG sample can be used as an SEY reference
sample, especially when baked. The SEY curve and the
$\delta_{max}$ obtained from the NEG were those expected, hence
ruling out any instrumental problem.

The last value for the $\delta_{max}$, reached after 40~mC/mm$^2$
of electron exposure was 2.1. Thus, the conclusion that we had reached
saturation, at the previous point, was hasty.


\subsection{Secondary Electron Yield of Al 6063}

\begin{figure}[htbp]
\centering
\includegraphics[width=0.6\textwidth,clip=]{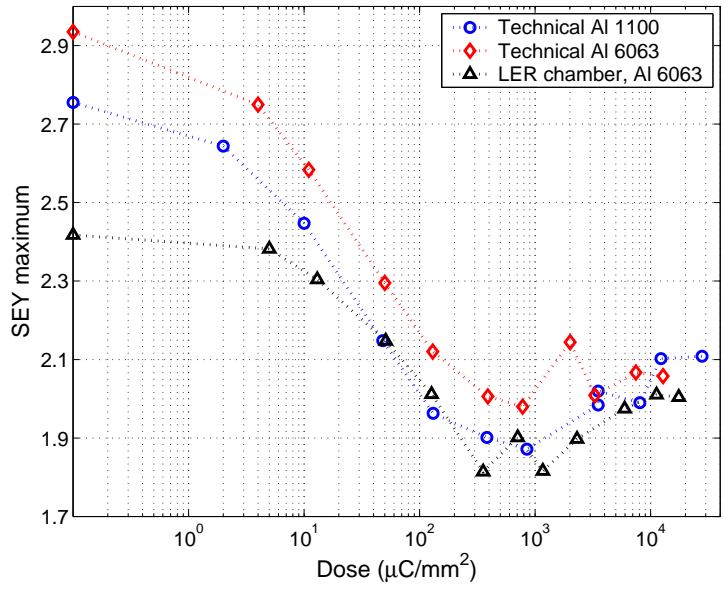}
\caption{SEY max vs the electron dose received by the Al alloy
samples, measured at 23$^\circ$ from normal incidence}
\label{figSEYvsdoseAl}
\end{figure}

\begin{figure}[htbp]
\begin{minipage}[t]{.5\linewidth}
\centering
\includegraphics[width=0.85\textwidth,clip=]{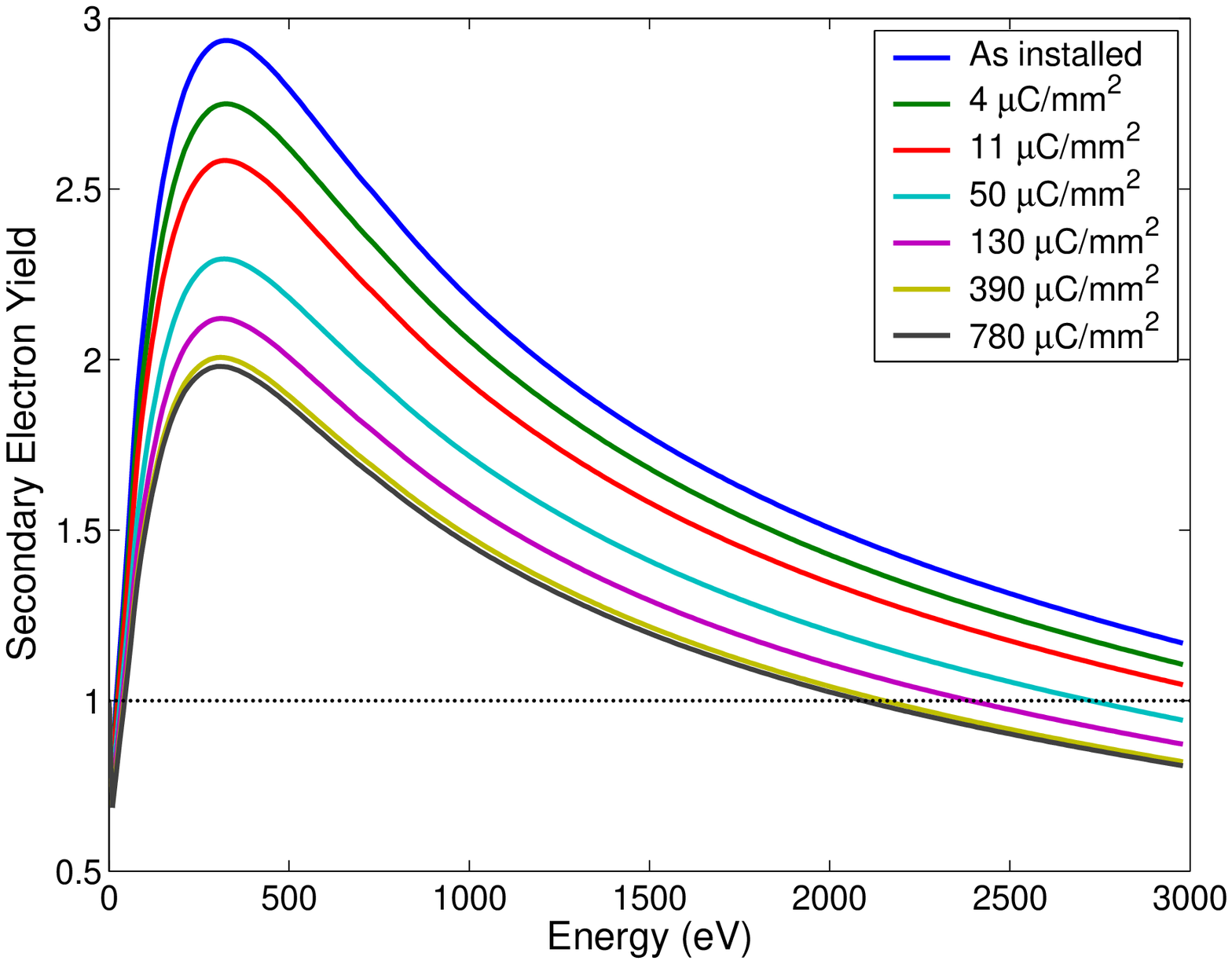}
\end{minipage}%
\begin{minipage}[t]{.5\linewidth}
\centering
\includegraphics[width=0.85\textwidth,clip=]{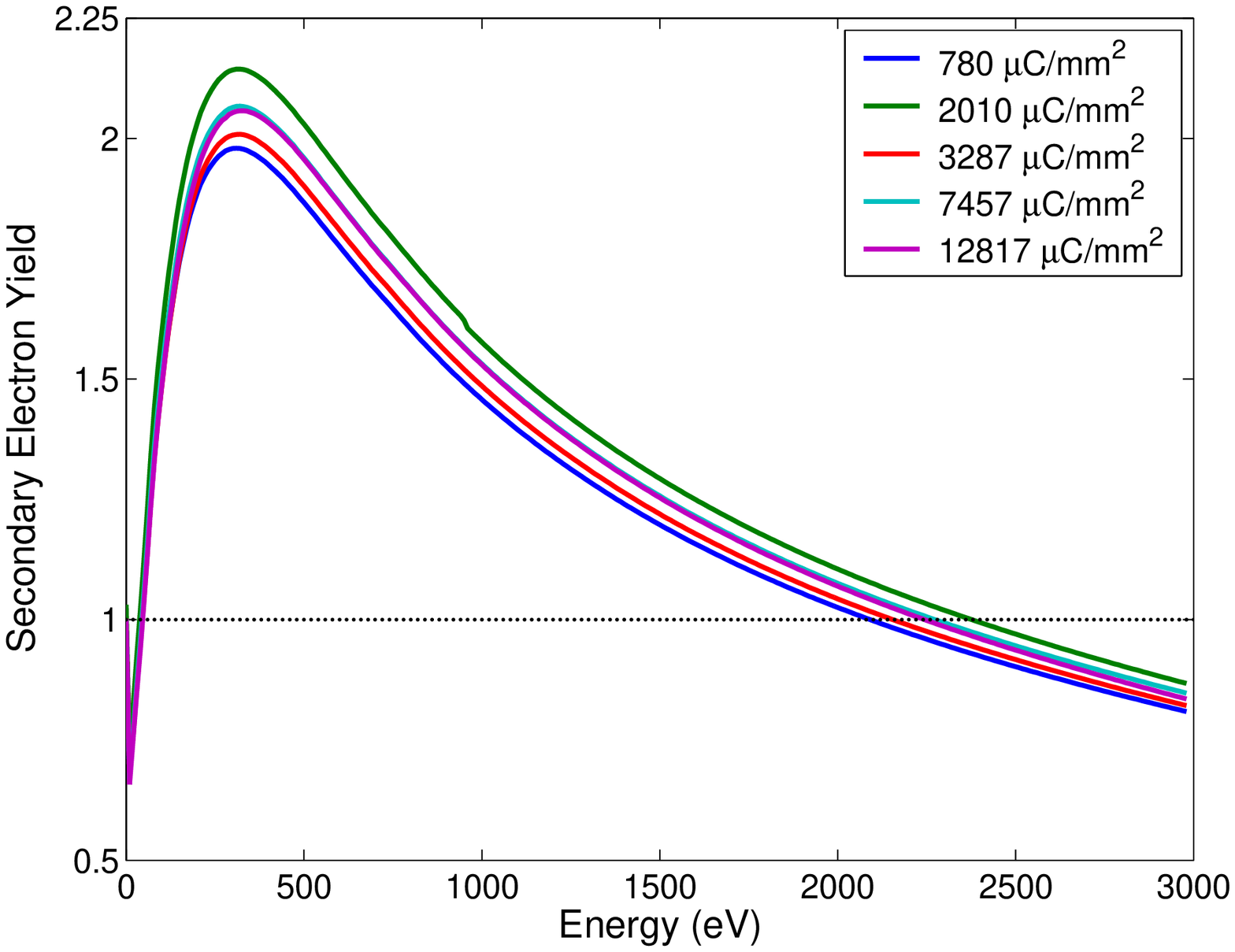}
\end{minipage}
\caption{Al 6063 exposed to electron conditioning. SEY values are
monotonically decreasing on the left plot and monotonically
increasing on the right plot} \label{figSEYal6063curves}
\end{figure}

\begin{figure}[htbp]
\begin{minipage}[t]{.5\linewidth}
\centering
\includegraphics[width=0.85\textwidth,clip=]{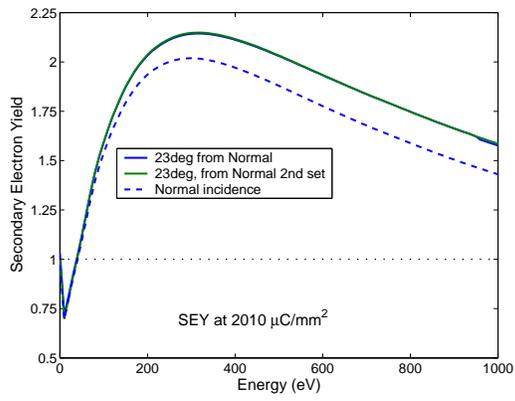}
\end{minipage}%
\begin{minipage}[t]{.5\linewidth}
\centering
\includegraphics[width=0.85\textwidth,clip=]{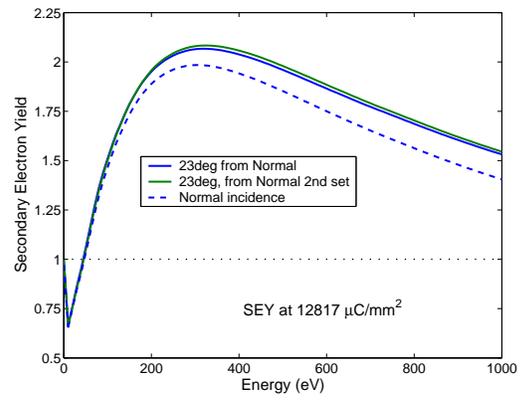}
\end{minipage}
\caption{SEY of Al 6063 at 23$^\circ$ and normal incidence}
\label{figSEYal6063angles}
\end{figure}

The SEY results obtained by exposing the Al 6063 sample to an
electron conditioning beam of 130~eV kinetic energy are presented
in Fig.\ref{figSEYvsdoseAl}, \ref{figSEYal6063curves} and
Fig.\ref{figSEYal6063angles}.

As observed for the Al 1100 sample, the SEY of the Al 6063 also
decreases with the increasing dose until reaching a dose of $\sim$
800~$\mu$C/mm$^2$, Fig.\ref{figSEYvsdoseAl} and
Fig.\ref{figSEYal6063curves}, left plot. After this point the SEY
increases, but not smoothly, Fig.\ref{figSEYvsdoseAl}. The SEY max
at a dose of 2010~$\mu$C/mm$^2$ reached a value of 2.13.
Subsequent measurements at this dose are in very good agreement
with the first set of data Fig.\ref{figSEYal6063angles}, left
plot. The next points at 3000~$\mu$C/mm$^2$, 7000~$\mu$C/mm$^2$
and 12000~$\mu$C/mm$^2$ have been also measured twice,
Fig.\ref{figSEYal6063angles} right plot as an example, and were
found to agree within 1.5\%. The SEY at those subsequent doses are
less than the one obtained at 2000~$\mu$C/mm$^2$,
Fig.\ref{figSEYvsdoseAl} and \ref{figSEYal6063curves}, right plot.
This jump is currently not understood.

\subsection{Secondary Electron Yield of the LER Al 6063}

\begin{figure}[htbp]
\centering
\includegraphics[width=0.6\textwidth,clip=]{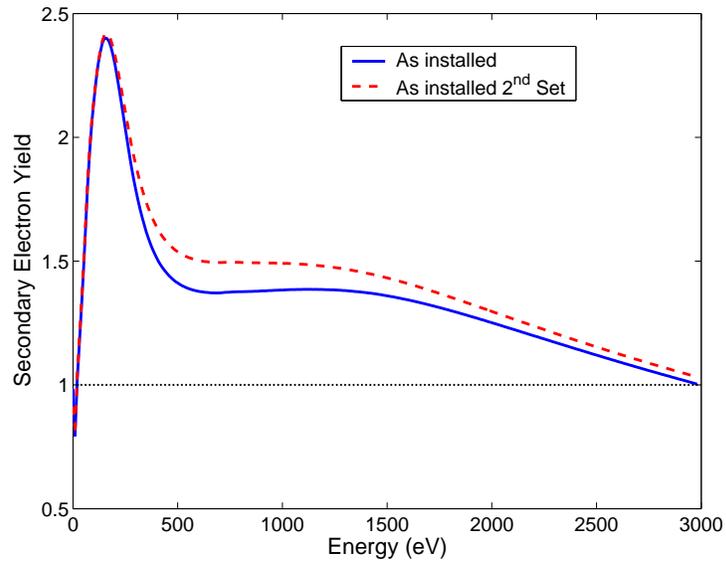}
\caption{SEY of LER Al 6063 as installed. Curve shapes are due to
charging of the oxide layer} \label{figSEYLERasinstalled}
\end{figure}

\begin{figure}[htbp]
\begin{minipage}[t]{.5\linewidth}
\centering
\includegraphics[width=0.85\textwidth,clip=]{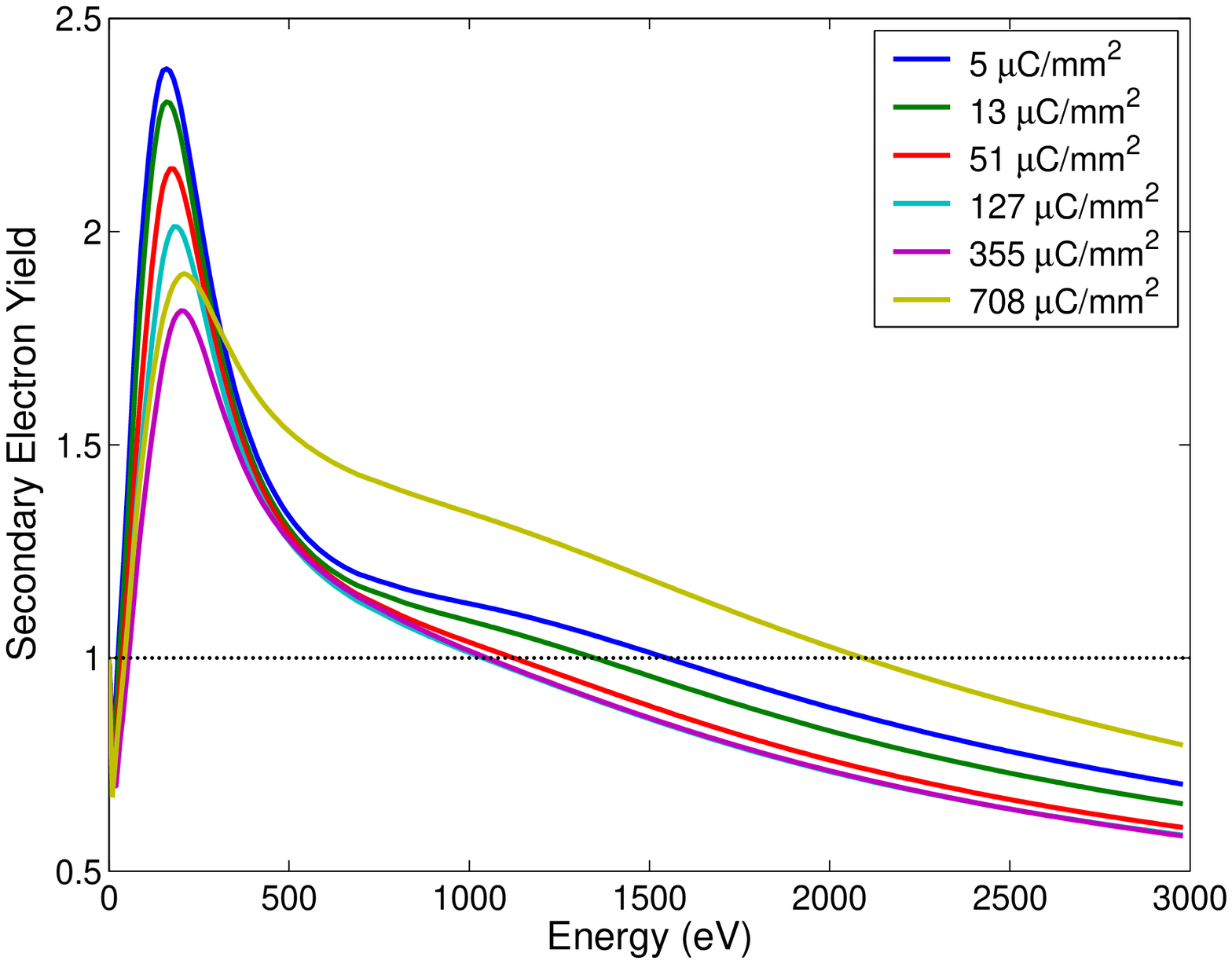}
\end{minipage}%
\begin{minipage}[t]{.5\linewidth}
\centering
\includegraphics[width=0.85\textwidth,clip=]{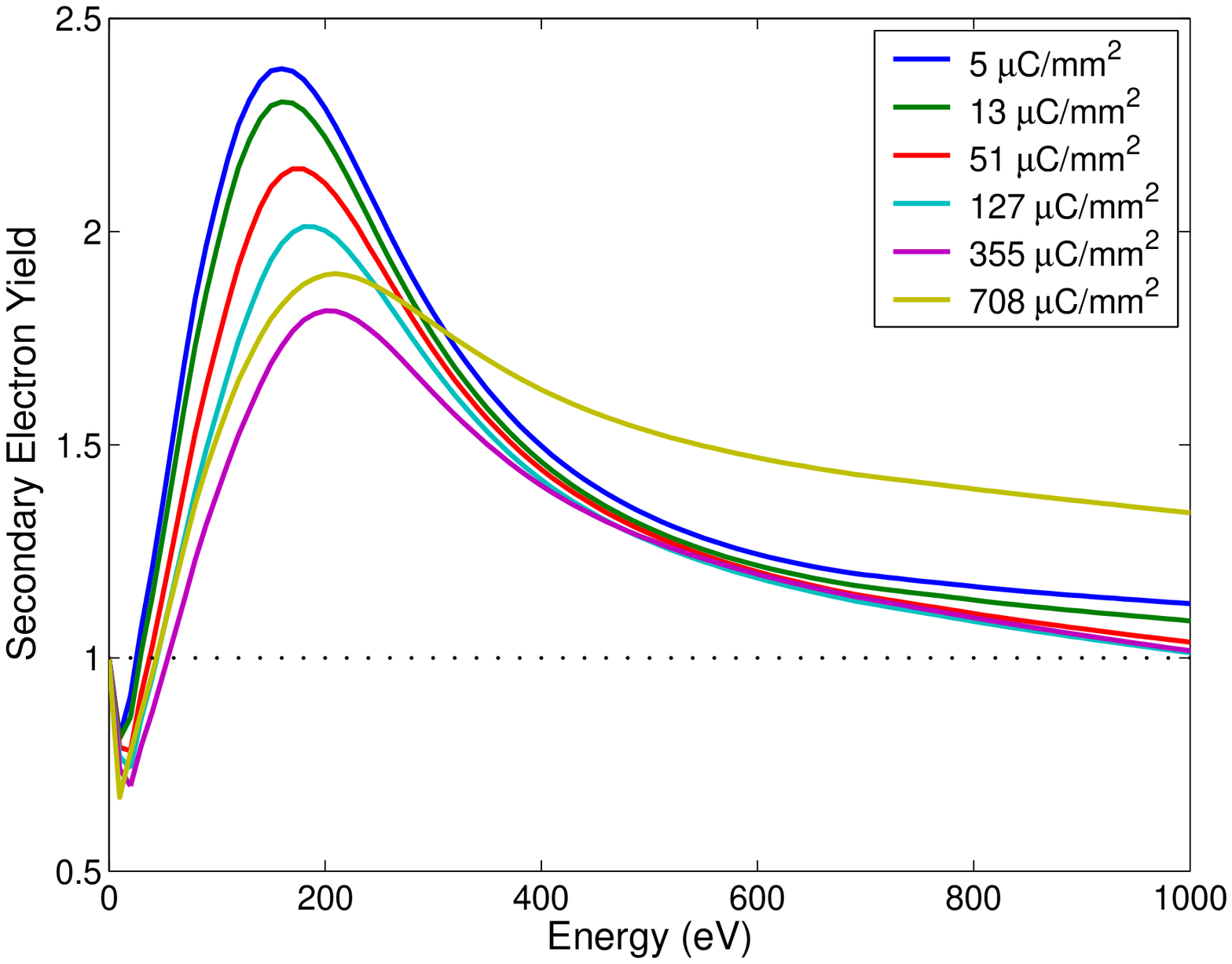}
\end{minipage}
\caption{LER Al 6063 exposed to electron conditioning, first part
of the conditioning. Values are monotonically decreasing. Detail
of the SEY between 0-1~keV right plots} \label{figSEYLER1part}
\end{figure}

\begin{figure}[htbp]
\begin{minipage}[t]{.5\linewidth}
\centering
\includegraphics[width=0.85\textwidth,clip=]{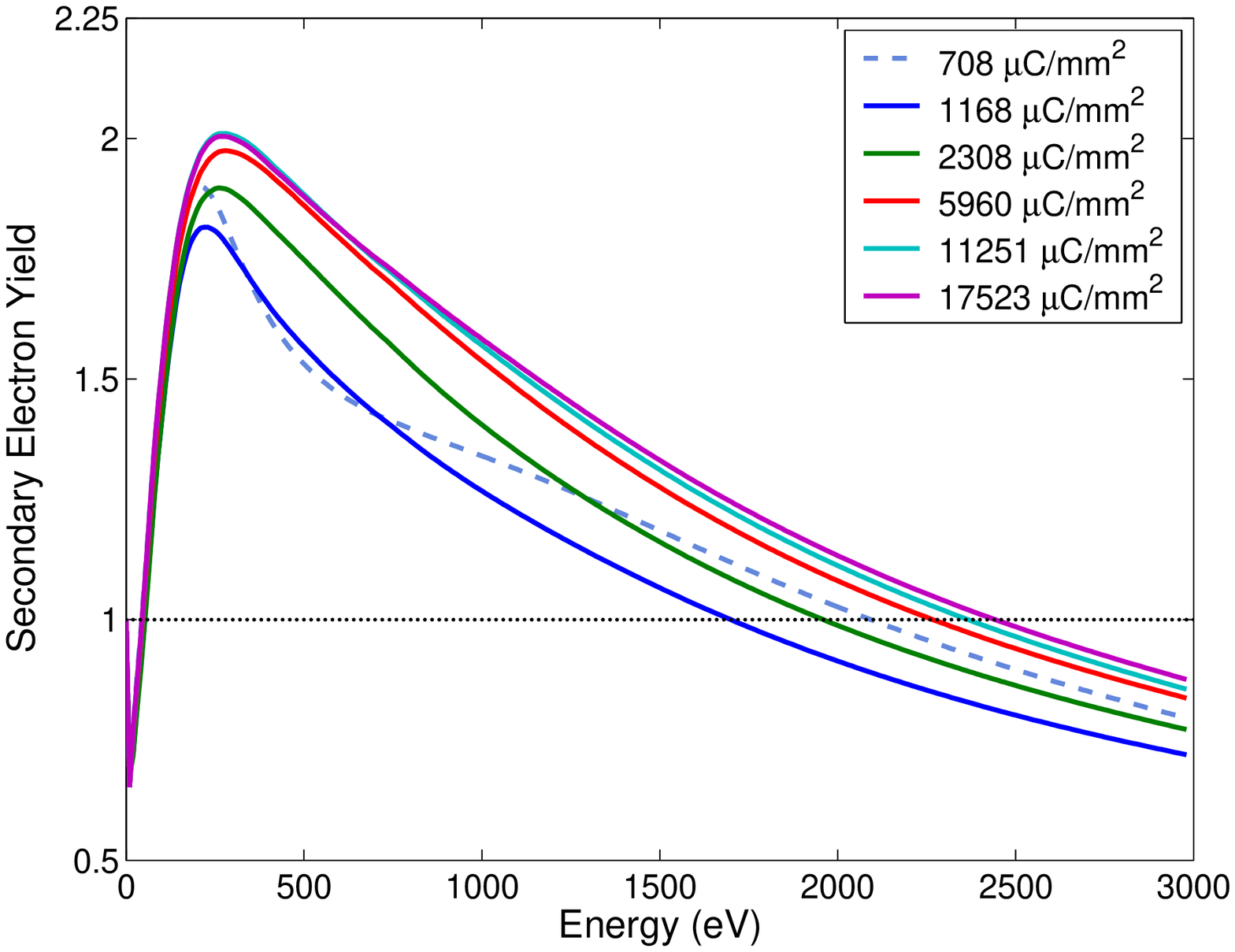}
\end{minipage}%
\begin{minipage}[t]{.5\linewidth}
\centering
\includegraphics[width=0.85\textwidth,clip=]{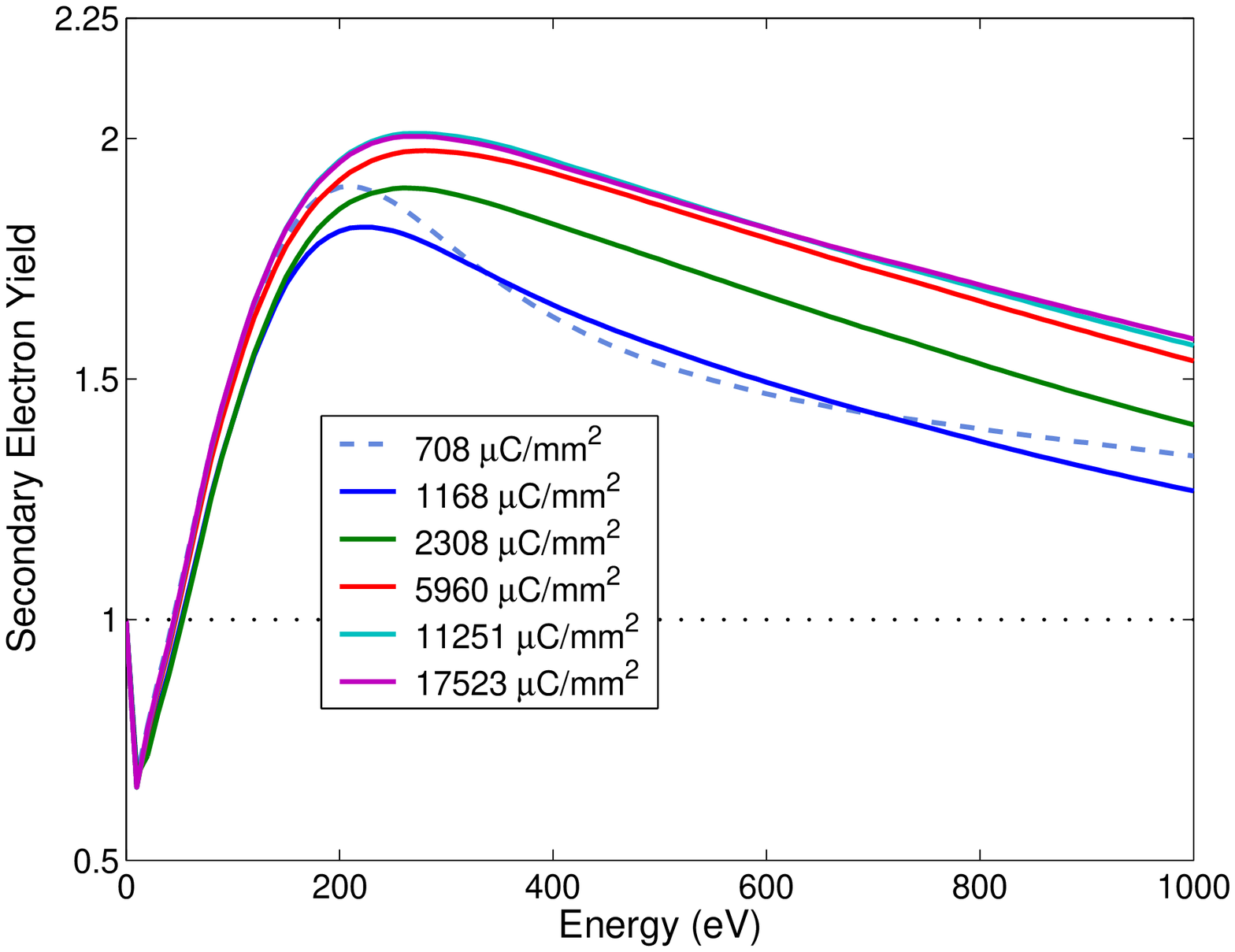}
\end{minipage}
\caption{LER Al 6063 exposed to electron conditioning, Second part
of the conditioning. Values are monotonically increasing.Detailed
of the SEY between 0-1~keV right plots} \label{figSEYLER2part}
\end{figure}

The SEY results obtained by exposing the LER Al 6063 sample to an
electron conditioning beam of 130~eV kinetic energy are presented
in Fig.\ref{figSEYLERasinstalled}, \ref{figSEYLER1part} and
Fig.\ref{figSEYLER2part}.

As observed for the Al 1100 and the Al 6063 scraped samples, the
SEY of the LER sample decreases with the increasing dose until
reaching a dose of $\sim$ 800~$\mu$C/mm$^2$,
Fig.\ref{figSEYvsdoseAl} and Fig.\ref{figSEYLER1part}, left plot.
After this point the SEY increases, but not smoothly,
Fig.\ref{figSEYvsdoseAl} and Fig.\ref{figSEYLER2part}. On the
second part of the dosing,Fig.\ref{figSEYLER2part}, the evolution
of the SEY changes drastically from the as-installed shapes and
values Fig.\ref{figSEYLERasinstalled}. The $\delta_{max}$ shifts
from being reached at 160~eV to being reached at 270~eV. The
curves after reaching a dose of 708~$\mu C/mm^2$ becomes similar
as the SEY plots from clean scraped Al 1100 and 6063.

\subsection{Are the results believable ?}

\begin{figure}[htbp]
\centering
\includegraphics[width=0.9\textwidth,clip=]{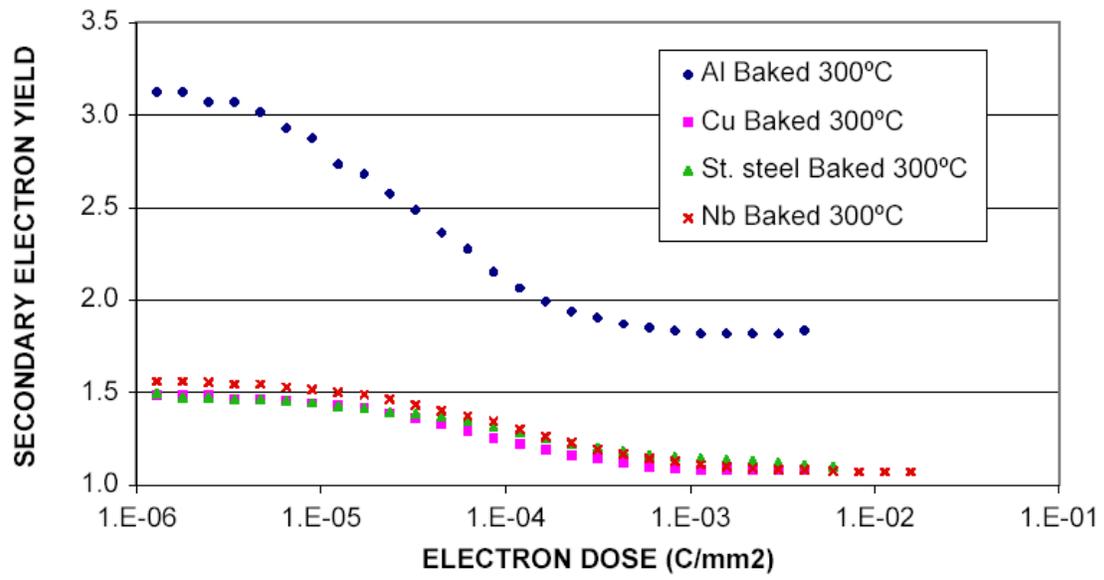}
\caption{SEY of baked technical surfaces conditioned by electrons
from ref.\cite{Hilleret:EPAC00}} \label{figSEYnoel}
\end{figure}

Despite the fact that our results seems contradictory to common
belief, previous data collected at CERN (Conseil Europ\'een pour
la Recherche Nucl\'eaire) \cite{Hilleret:EPAC00} supports,
indirectly, our findings. The conditioning curves,
Fig.\ref{figSEYnoel}, of the 300$^\circ$C pre-baked aluminium
sample (blue circle) show a dip, which bottoms around 1.8, data
taken at normal incidence. The SEY of the last aluminium point,
around 7~mC/mm$^2$, is in very good agreement with our value
obtained at 8~mC/mm$^2$, Fig.\ref{figSEYvsdoseAl}.

In some other data, collected at ANL (Argonne National Laboratory)
on an Al 6063 sample, the $\delta_{max}$ achieved after an
electron dose exposure of 350~nA/cm$^2$ for 5h (equivalent to
63~$\mu$C/mm$^2$) at an energy of 100~eV is around 2.1
\cite{Rosenberg:2003}. From these comparisons and the system
check, by the use of a NEG, we are confident in the results
obtained above 8~mC/mm$^2$.

\section{XPS study of the C1s and Al2p peak}

\subsection {XPS study of Al 1100}

\begin{figure}[htbp]
\centering
\includegraphics[width=0.8\textwidth,clip=]{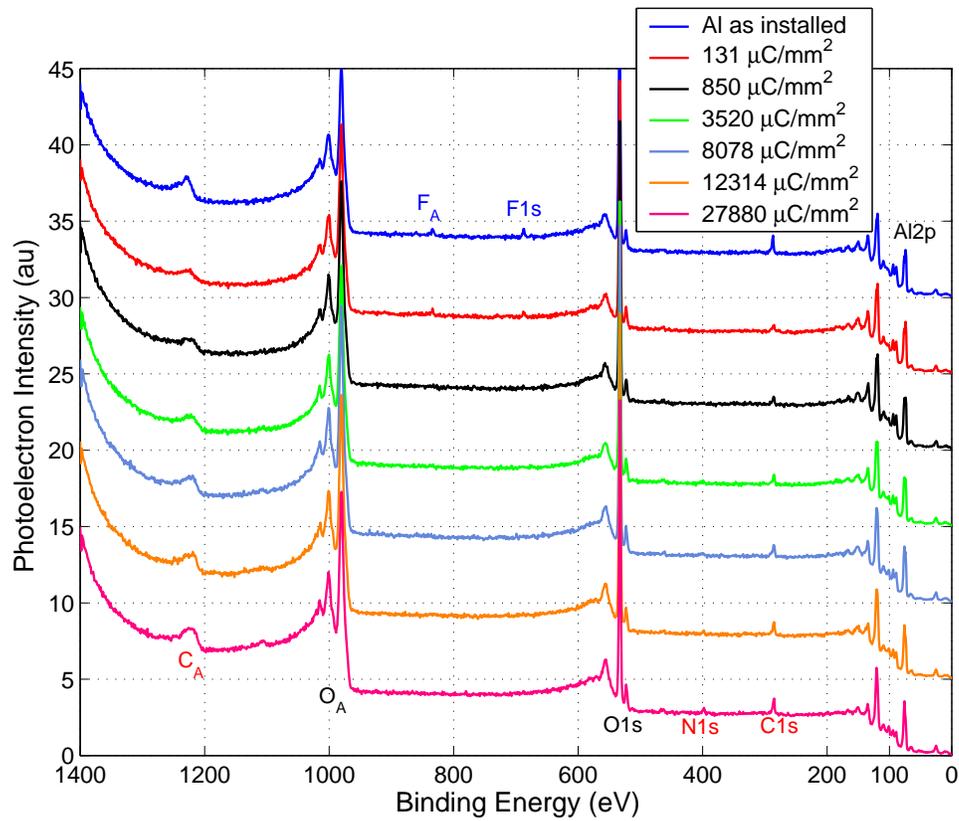}
\caption{XPS survey of the Al 1100 sample during electron
conditioning. "A" subscript indicates Auger peak. Top curve: as
installed sample; near BE axis: sample after 27880~$\mu$C/mm$^2$
electron dosing} \label{figXPSsurveyAl}
\end{figure}

XPS analysis was carried out to observe the evolution of the
carbon and aluminium chemistry during the electron conditioning,
Fig.\ref{figXPSsurveyAl}. The spectra are shifted vertically from
one another for clarity. The "Al as installed" spectra is located
at the top of Fig.\ref{figXPSsurveyAl}, and the "27880
$\mu$C/mm$^2$" at its bottom.

\begin{figure}[tbp]
\centering
\includegraphics[width=0.92\textwidth,clip=]{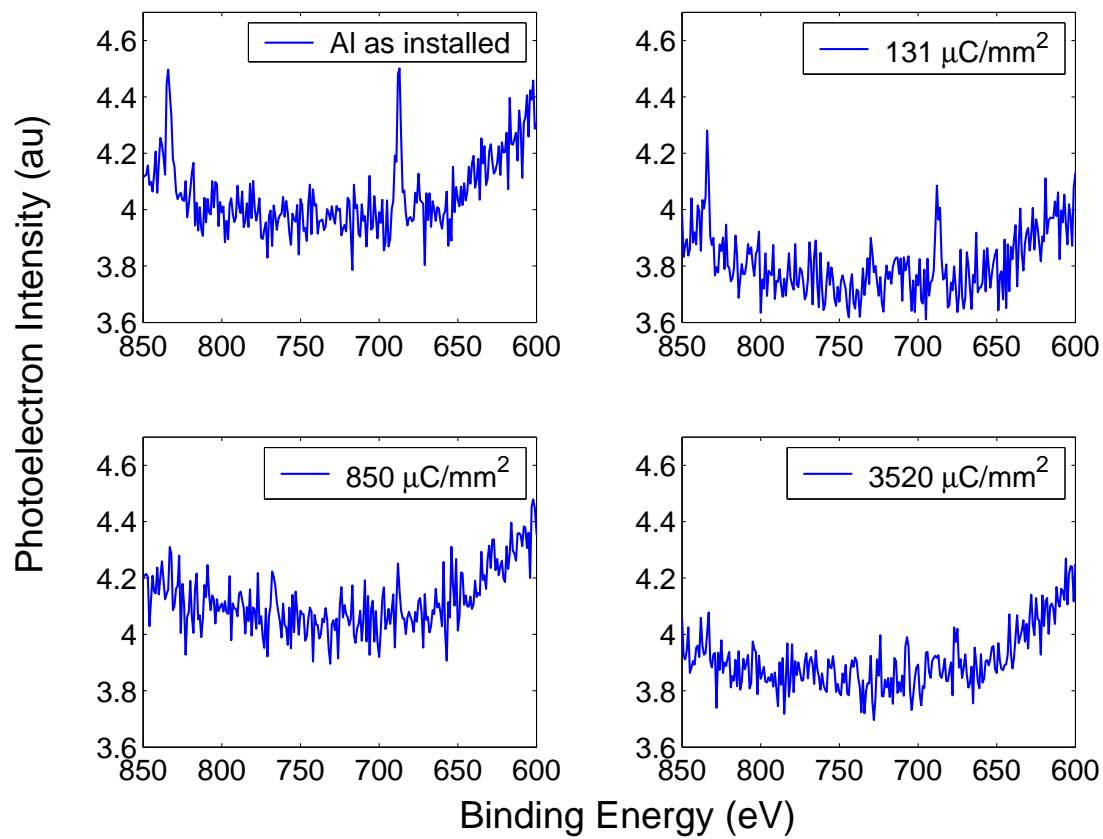}
\caption{Disappearance of F1s (685~eV)and
F$_A$ (Auger) during initial electron conditioning}
\label{figXPSfluor}
\end{figure}

\begin{figure}[tbp]
\centering
\includegraphics[width=0.92\textwidth,clip=]{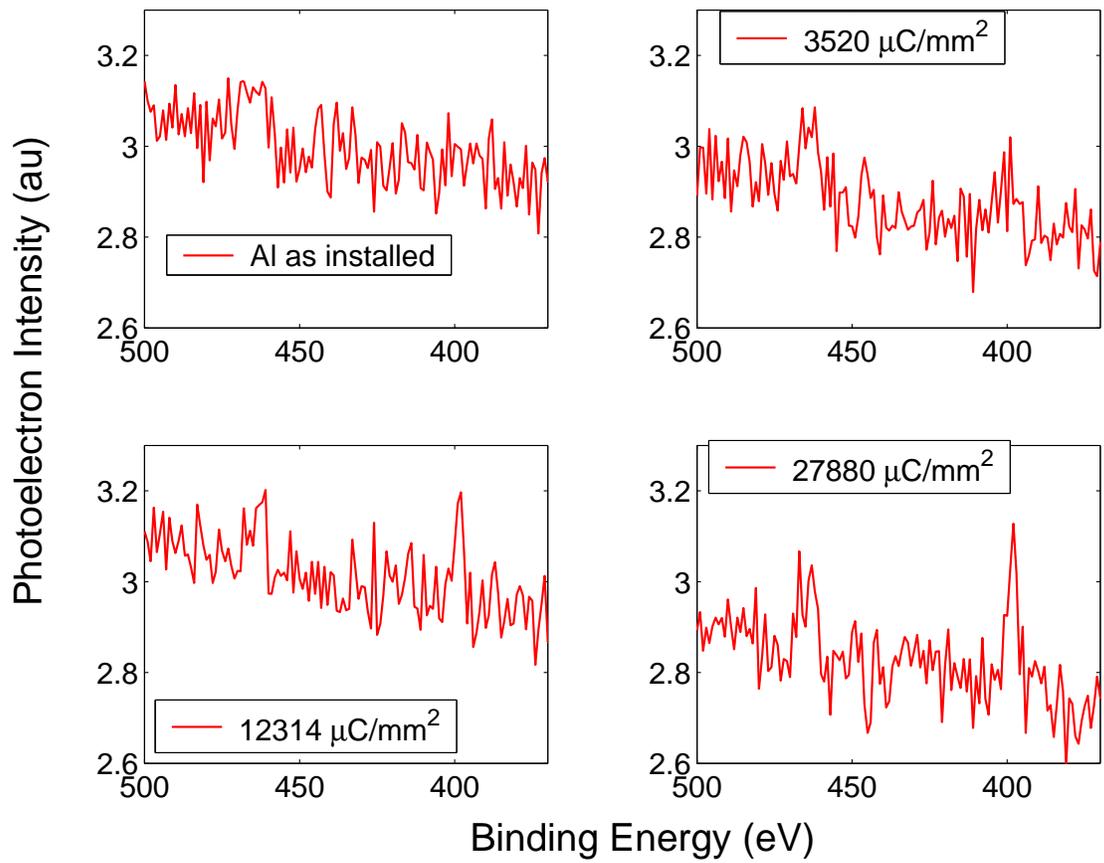}
\caption{Emergence of N1s (400~eV) during
late electron conditioning} \label{figXPSazote}
\end{figure}

From the "as installed" condition to the end of the conditioning, a
few obvious observation can be done. First of all, the "as
installed" sample is contaminated by fluorine (F). The sample was
not passivated and was thoroughly scraped. We must assume that
this F is present in the air and reacts very quickly with a pure
Al surface, hence getting imbedded in the oxide layer. Fluorine
compounds are used heavily in the semiconductor industry to
prepare silicon wafers. Our location is in the heart of this
industry.
\newline During the initial conditioning the F1s (685~eV) quickly disappears,
Fig.\ref{figXPSfluor}. However, a peak of nitrogen then appears,
N1s (398~eV), Fig.\ref{figXPSazote}.
\newline It is possible that during our attempt to create an AlN
film, a proportion of N was absorbed in the bulk of the Al 1100.
During conditioning, the surface is "cleaned-up", by ESD and the
mobility of the N is enhanced, thence diffusing to the surface or
near subsurface(1-5 nm depth). This N concentration, being very
small, 2~at\%, is unlikely to have influenced the behaviour of the
aluminium with respect to the SEY.

\begin{figure}[htbp]
\centering
\includegraphics[width=0.7\textwidth,clip=]{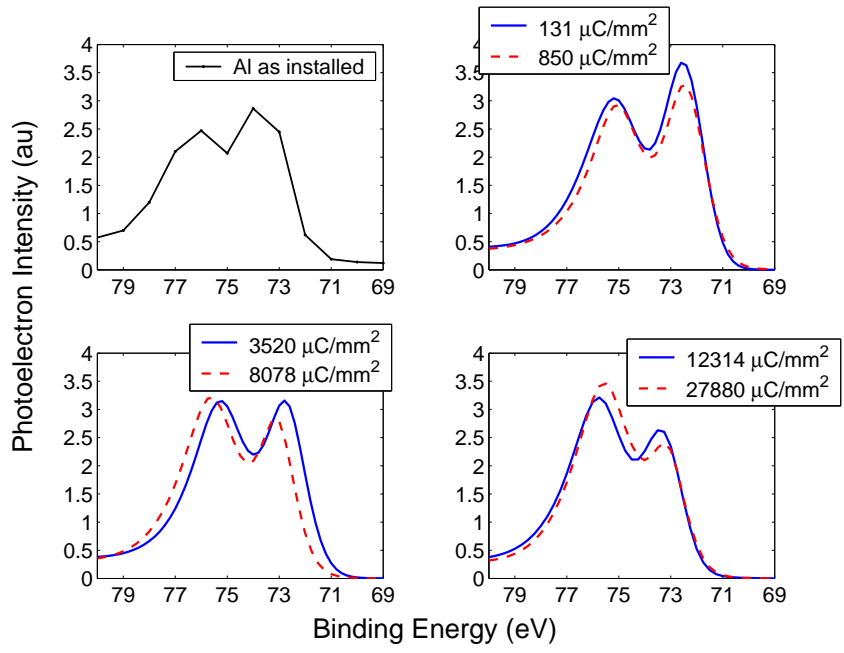}
\caption{Detailed spectra of the Al2p during the electron
conditioning} \label{figXPSWAl}
\end{figure}

A survey of the Al2p was carried out during the conditioning. The
high-energy resolution spectra are shown in Fig.\ref{figXPSWAl}. A
pure Al surface will present a single peak at 73~eV, and a pure
Al$_2$O$_3$ surface will have one peak at 74.5~eV
\cite{XPShandbook:92}.

The "as installed" Al2p is peaked at 73.5~eV and 76.5~eV, shown at
a resolution of 0.5~eV for a step scan energy of 1~eV. Those peaks
match the Al2p location of a pure Al surface and a halogenated Al
surface. As the fluorine disappears from the surface, the spectrum
shifts to lower binding energy and presents the characteristic of
a thin aluminium oxide film (less than 5 nm) on an aluminium
substrate. Very similar curves on an Al 6063 alloy sample can be
found in \cite{Rosenberg:2003}, where the peaks are
representatives of the pure Al (73~eV BE) and the oxidized Al
(76~eV BE). Moreover, during conditioning of our Al 1100 sample,
the relative intensities of the peaks changes during the
conditioning. The aluminium peak (73~eV) becomes smaller than the
oxide peak (75~eV).\newline From this last observation we
hypothesize that we are thickening the aluminium oxide layer by
decomposing carbon monoxide and dioxide from the residual gas and,
by rearranging the bonds on the surface, the oxygen displaces the
carbon covering the aluminium.

\begin{figure}[htbp]
\centering
\includegraphics[width=0.7\textwidth,clip=]{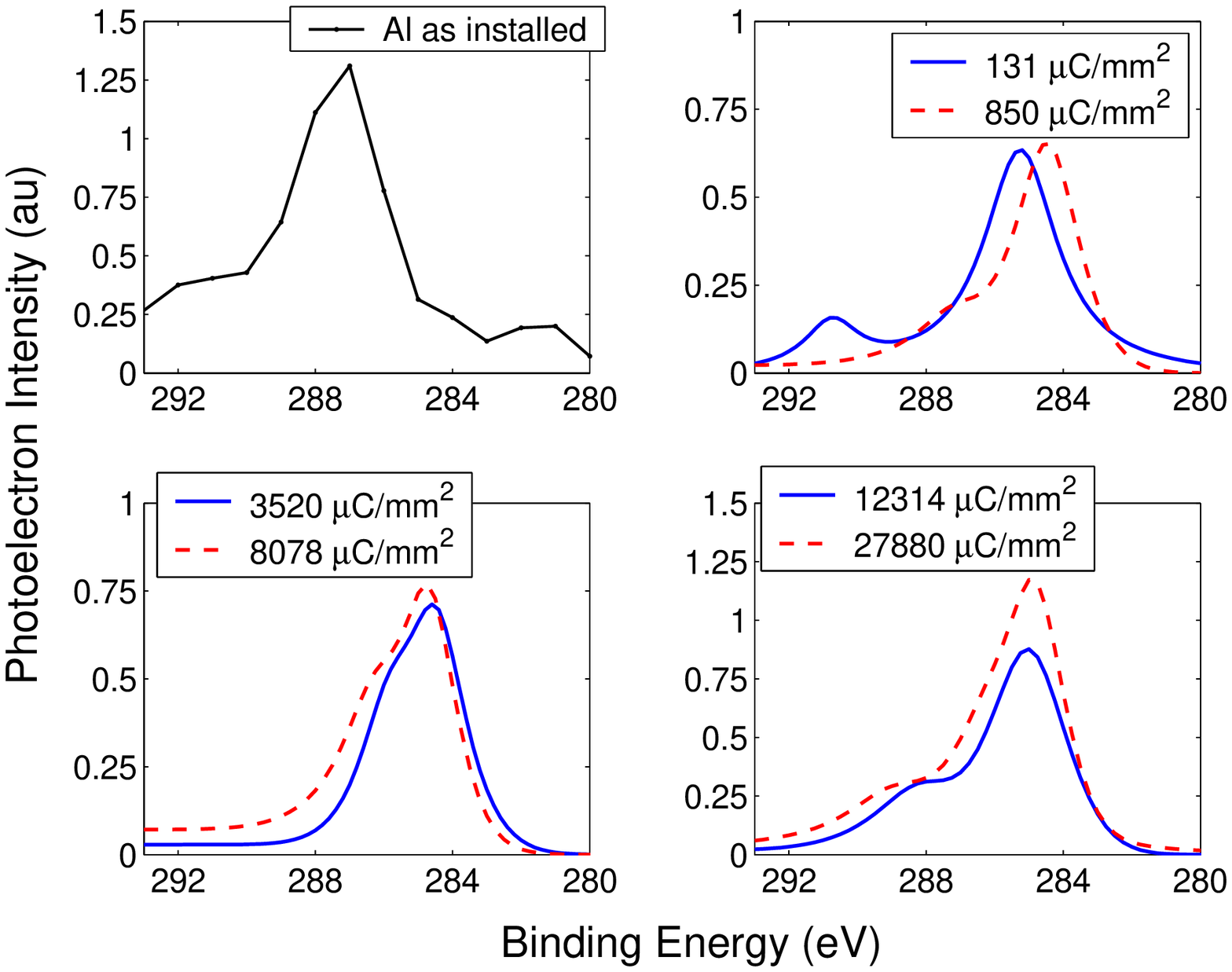}
\caption{Detailed spectrum of the C1s during electron
conditioning} \label{figXPSWC}
\end{figure}

This interpretation for the Al is supported by the C1s spectra,
Fig.\ref{figXPSWC}. The C of the "as installed" Al is peaked at
287.5~eV, but also has a peak at 291~eV. This high binding energy
(BE) is reminiscent of carbon passivated by HF acid which shows a
peak at 289~eV \cite{XPShandbook:92}. CF$_2$ compounds also will
have a peak at 292~eV \cite{XPShandbook:92}. After an accumulated
dose of 850~$\mu$C/mm$^2$, the F disappeared, and we saw a shift
from 291~eV to 288~eV, location of oxidized C1s. During
conditioning, the peak not only gets shifted further toward 285~eV
(marker of an amorphous/graphitic C surface) but also the peak
intensity rose. This shows that the C is transformed from an
oxidized state to its amorphous/graphitic form.

\subsection{XPS of Al 6063}

\begin{figure}[htbp]
\centering
\includegraphics[width=0.8\textwidth,clip=]{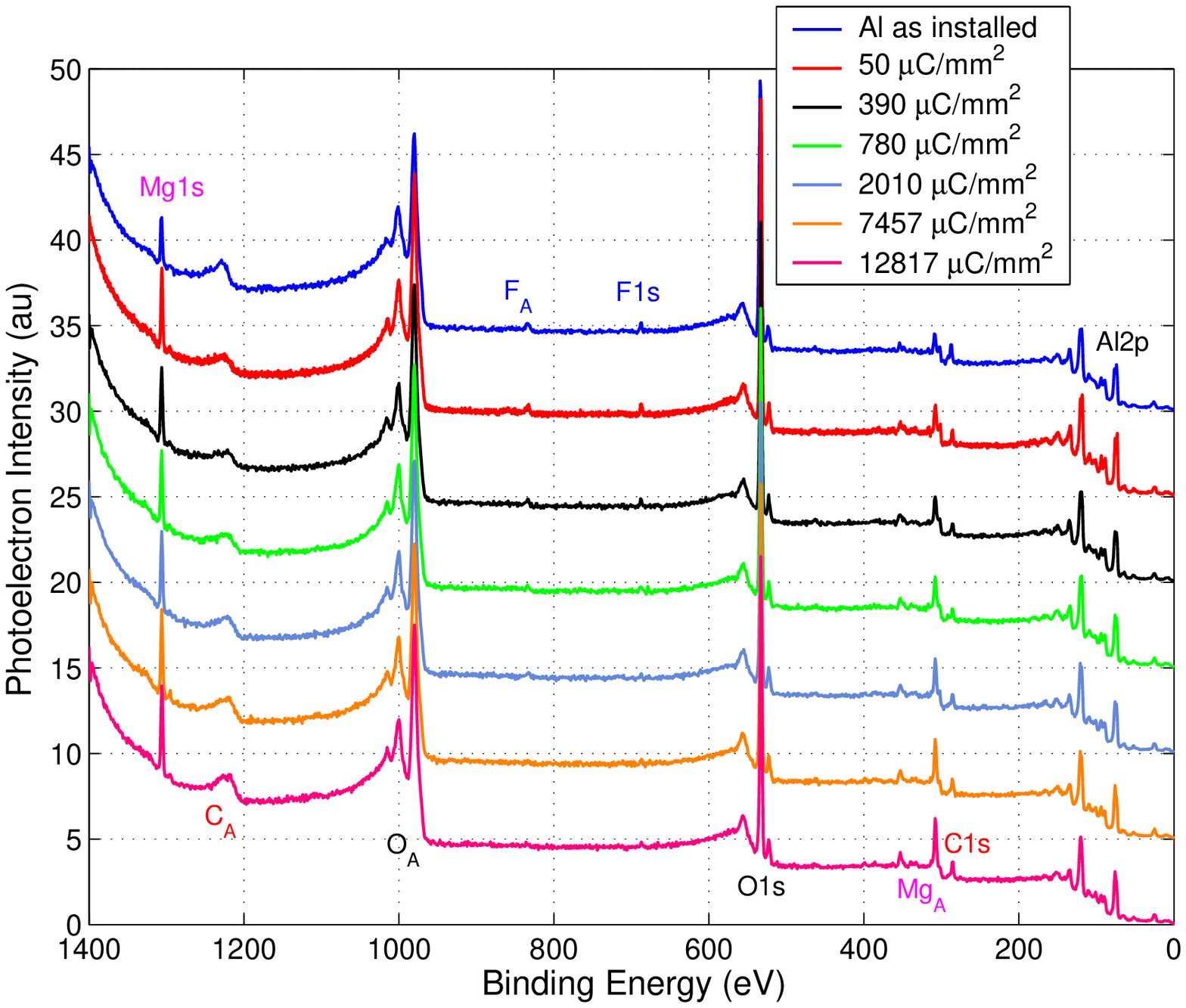}
\caption{XPS survey of Al 6063 sample during electron
conditioning. Top curve: as installed sample; near BE axis: sample
after 12817~$\mu$C/mm$^2$ electron dosing}
\label{figXPSsurveyAl6063}
\end{figure}

XPS analysis of Al 6063 was carried out to observe the evolution
of the carbon and aluminium chemistry during the electron
conditioning, Fig.\ref{figXPSsurveyAl6063}. The spectra are
shifted vertically from one another for clarity. The "Al as
installed" spectra is located at the top of
Fig.\ref{figXPSsurveyAl6063}, and the "12817 $\mu$C/mm$^2$" near
the BE axis.

\begin{figure}[htbp]
\centering
\includegraphics[width=0.7\textwidth,clip=]{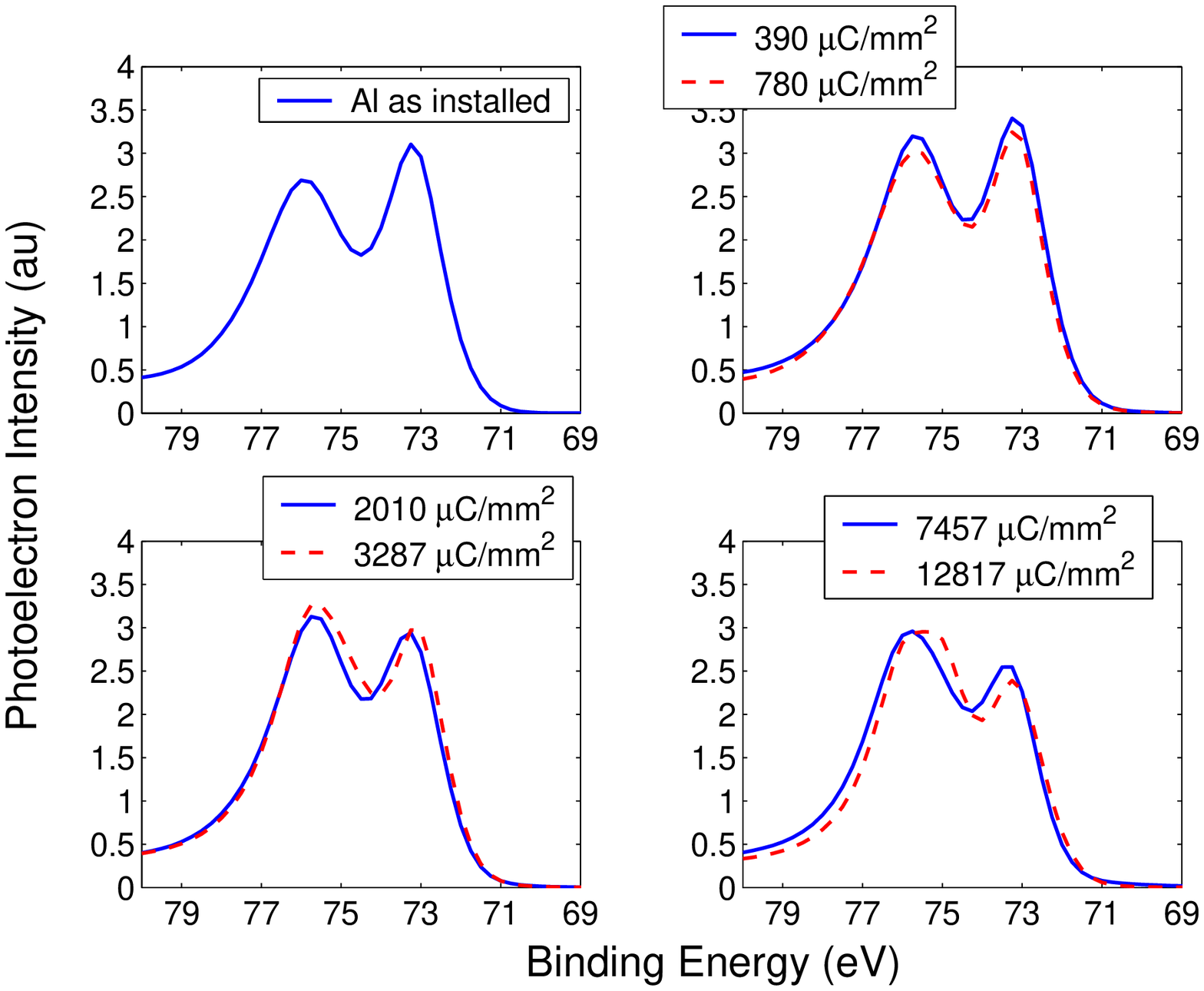}
\caption{Detailed spectra of the Al 2p during electron
conditioning of Al 6063} \label{figXPSWAl6063}
\end{figure}

\begin{figure}[tbp]
\centering
\includegraphics[width=0.92\textwidth,clip=]{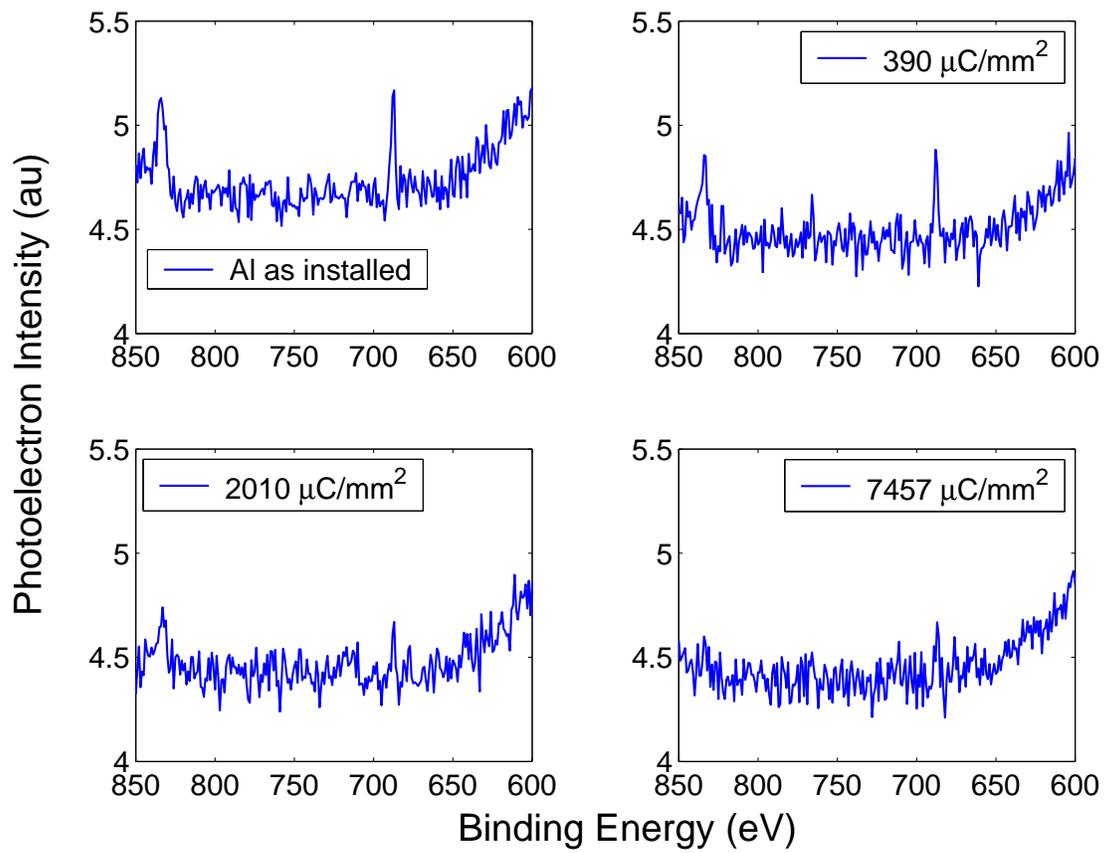}
\caption{Disappearance of F1s (685~eV)
during electron conditioning} \label{figXPSfluor6063}
\end{figure}

The observations on Al 6063 are similar of those on Al 1100. The
"as installed" Al2p is peaked at 73.5~eV and 76~eV, for a step
scan energy of 0.25~eV, Fig.\ref{figXPSWAl6063}. Those peaks match
the location of a pure Al surface and a halogenated Al surface. As
the fluorine disappears from the surface, due to electron
bombardment, the spectrum shifts to lower binding energy and
presents the characteristic of a thin aluminium oxide film (less
than 5~nm) on an aluminium substrate, Fig.\ref{figXPSfluor6063}.
The shift in energy, is also accompanied with a change in
intensities between the peaks, as it was observed on the Al 1100,
Fig.\ref{figXPSWAl}. In \cite{Rosenberg:2003}, the XPS spectra
from the technical Al 6063 sample, exposed to the effect of a
running accelerator beam, shows a shift toward lower BE of the
oxidized peak. The intensity of the oxidized Al peak before
running the beam is higher than the pure Al peak. It does not
seems that the relative intensities between the pure metal and
oxidized metal varies at the end of the beam exposure. The XPS
spectra before and after exposure to the beam of the facing away
side of the sample shows no variation.

\begin{figure}[tbp]
\centering
\includegraphics[width=0.92\textwidth,clip=]{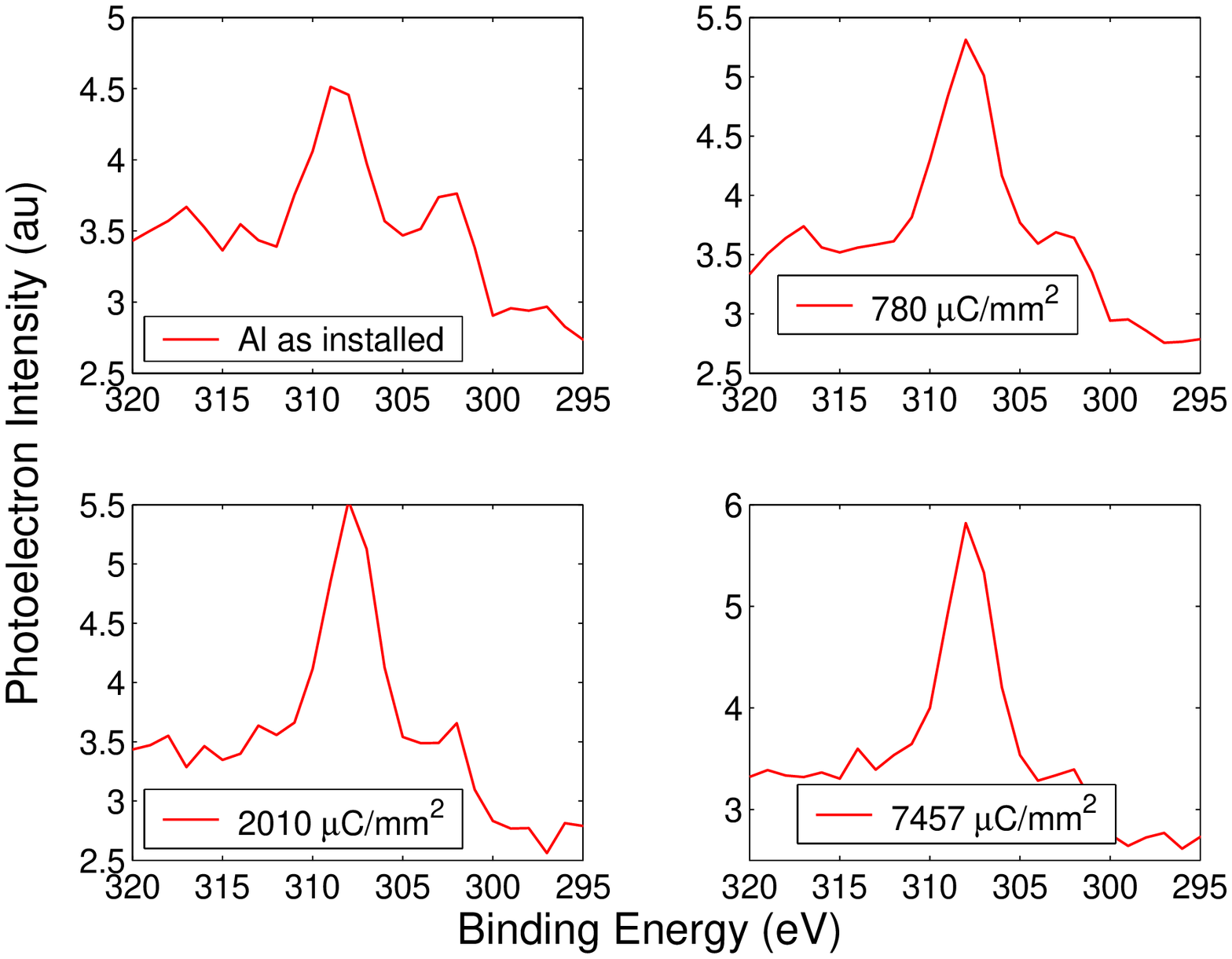}
\caption{Modification of Auger
KL$_{23}$L$_{23}$ Mg peaks (301~eV and 308~eV) during electron
conditioning} \label{figXPSMg6063}
\end{figure}

The presence of significant amounts of Mg inside the 6063 alloy
complicates the interpretation of the XPS data. Its continuous
presence on the surface is marked by its 1s peak at 1306~eV BE,
Fig.\ref{figXPSsurveyAl6063}. Mg is also a very good oxygen getter
and its evolution was monitored by observing the 301~eV and 308~eV
BE KLL Auger peaks, Fig.\ref{figXPSMg6063}. A pure Mg surface will
present an higher Auger peak at 301~eV that at 308~eV,
\cite{XPShandbook:92}; this picture is reversed for an oxidized Mg
\cite{Asami:2000}. During the conditioning we see that the 308~eV
peaks increases and the 301~eV disappears.

An XPS spectrum from a piece of LER vacuum chamber, made of Al
6063, was also taken. The spectrum does not show any pure metal
peak at 301~eV. This piece of LER chamber was kept in air for many
years, hence built a thick natural Al and Mg oxide, probably
Mg(OH)$_2$ \cite{Asami:2000}. All of these observations support
our preceding hypothesis, in which we stated that, during electron
conditioning an oxide layer grows on the technical surface.

\begin{figure}[htbp]
\centering
\includegraphics[width=0.7\textwidth,clip=]{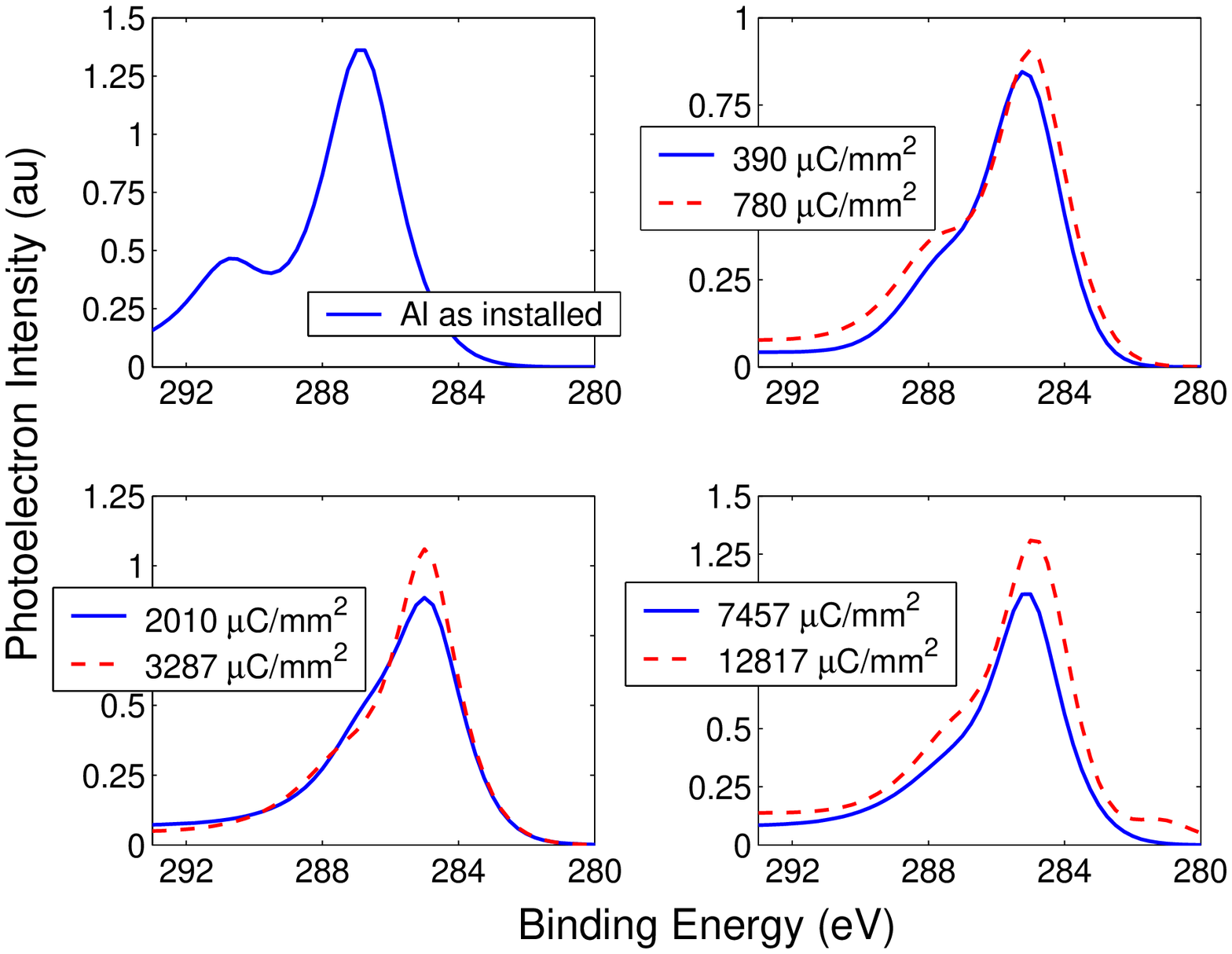}
\caption{Detailed spectrum of the C1s during electron conditioning
of Al 6063} \label{figXPSWC6063}
\end{figure}

The Al 6063 XPS spectrum for the C1s (285~ev BE),
Fig.\ref{figXPSWC6063}, is similar to the one obtained for the Al
1100, Fig.\ref{figXPSWC}. The results obtained on the two alloy
surfaces show the same chemistry evolution.

\subsection{XPS of LER Al 6063}

\begin{figure}[htbp]
\centering
\includegraphics[width=0.8\textwidth,clip=]{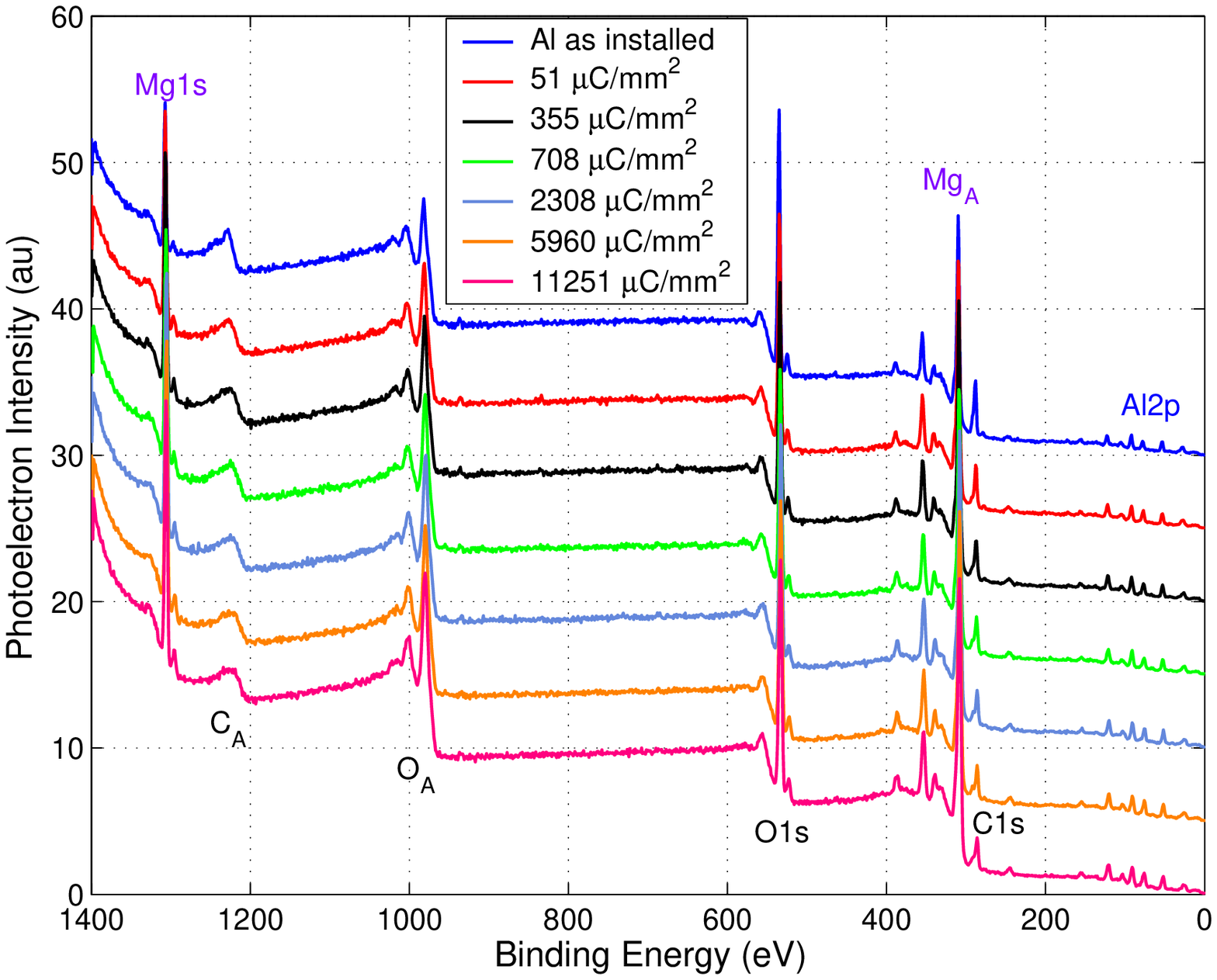}
\caption{XPS survey of LER (Al 6063) sample during electron
conditioning. Top curve: as installed sample; near BE axis: sample
after 11251~$\mu$C/mm$^2$ electron dosing} \label{figXPSsurveyLER}
\end{figure}

XPS analysis of LER Al 6063 was carried out to observe the
evolution of the carbon and aluminium chemistry during the
electron conditioning, Fig.\ref{figXPSsurveyLER}. The spectra are
shifted vertically from one another for clarity. The "Al as
installed" spectra is located at the top of
Fig.\ref{figXPSsurveyLER}, and the "12817~$\mu$C/mm$^2$" near the
BE axis. The survey shows that a non scraped Al 6063 has a
different chemistry at the surface than the scraped Al 6063
Fig.\ref{figXPSsurveyAl6063}. There is, for example, no presence
of fluorine (685~eV) on the surface.

The Al2p spectra of the non scraped Al 6063 sample coming from the
LER chamber shows the presence of one peak at 78~eV. A non
oxidized Al surface has an Al2p peak, peaked at 73~eV
\cite{XPShandbook:99}. As no fluorine is present, this peak could
be the mark of an oxide. However, as seen in \cite{XPShandbook:92}
Al$_2$O$_3$ is peaked at 74.5~eV, however from
\cite{XPShandbook:99}, Al$_2$O$_3$ is peaked at 75.6~eV and
Al(OH)$_3$ at 76.5~eV. None of the oxide explain the shift to the
higher (78~eV) BE. However, as the electron dose received by the
sample increased, the peak shifts to lower BE and broadens. This
broadening is the signature of the rearrangement of the aluminium
oxide.

\begin{figure}[tbp]
\centering
\includegraphics[width=0.92\textwidth,clip=]{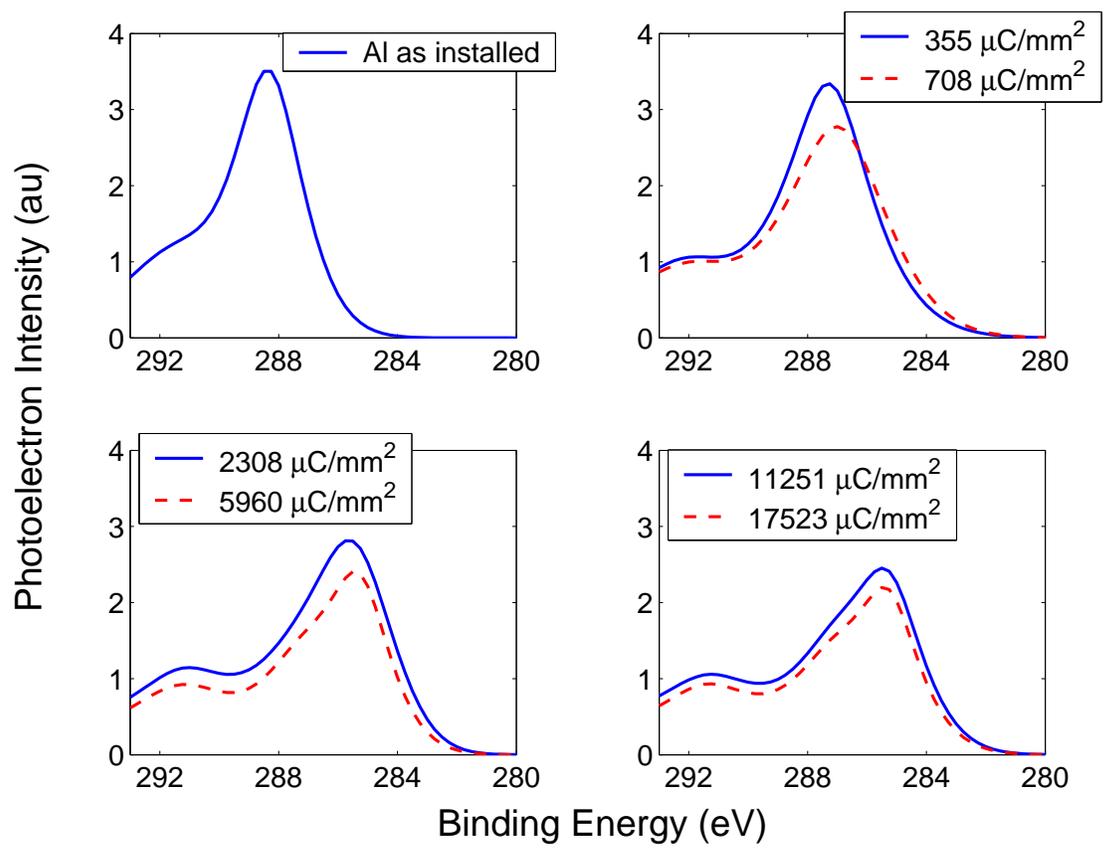}
\caption{Evolution of C1s during electron
conditioning of the LER Al 6063} \label{figXPSWCLER6063}
\end{figure}

The LER Al 6063 XPS spectrum for the C1s (285~ev BE),
Fig.\ref{figXPSWCLER6063}, is similar to the one obtained for the
Al 1100 and scraped Al 6063, Fig.\ref{figXPSWC} and
Fig.\ref{figXPSWC6063}. Their evolution during electron
bombardment are also similar, the oxidized carbon being
transformed into a more graphitic form. It was hypothesized that
the C peak at 291~eV for the scraped Al technical surfaces was due
to the presence of F at the surface. As no F is present on the LER
Al 6063, the deformed shape of the C must be due to different
oxygen bonding. Oxide containing hydroxide-type (C-O-H) produces a
chemical shift of 1.5~eV from the graphitic carbon (285~eV BE).
Carbonyl (C=O) and carboxyl (COOH) have chemical shift of 3.0~eV
and 4.5~eV, respectively \cite{Sherwood:1994}. As this sample was
kept in air for years, all oxidation state of the surface are
likely. The intensity of the "as installed" C1s peak of the LER
6063 sample, Fig.\ref{figXPSsurveyLER}, is higher than that for
the scraped 6063 and 1100 samples, Fig.\ref{figXPSsurveyAl6063}
and Fig.\ref{figXPSsurveyAl}, respectively. A fair amount of
carbonate (CO$_3$) can also be present, with the BE of this
compound on a native Al oxide peaked at 291~eV
\cite{XPShandbook:99}. However, it is stated in
\cite{Buckley:2003} that aluminium carbonate is not expected to be
formed in presence of CO$_2$ and water.

\begin{figure}[tbp]
\centering
\includegraphics[width=0.92\textwidth,clip=]{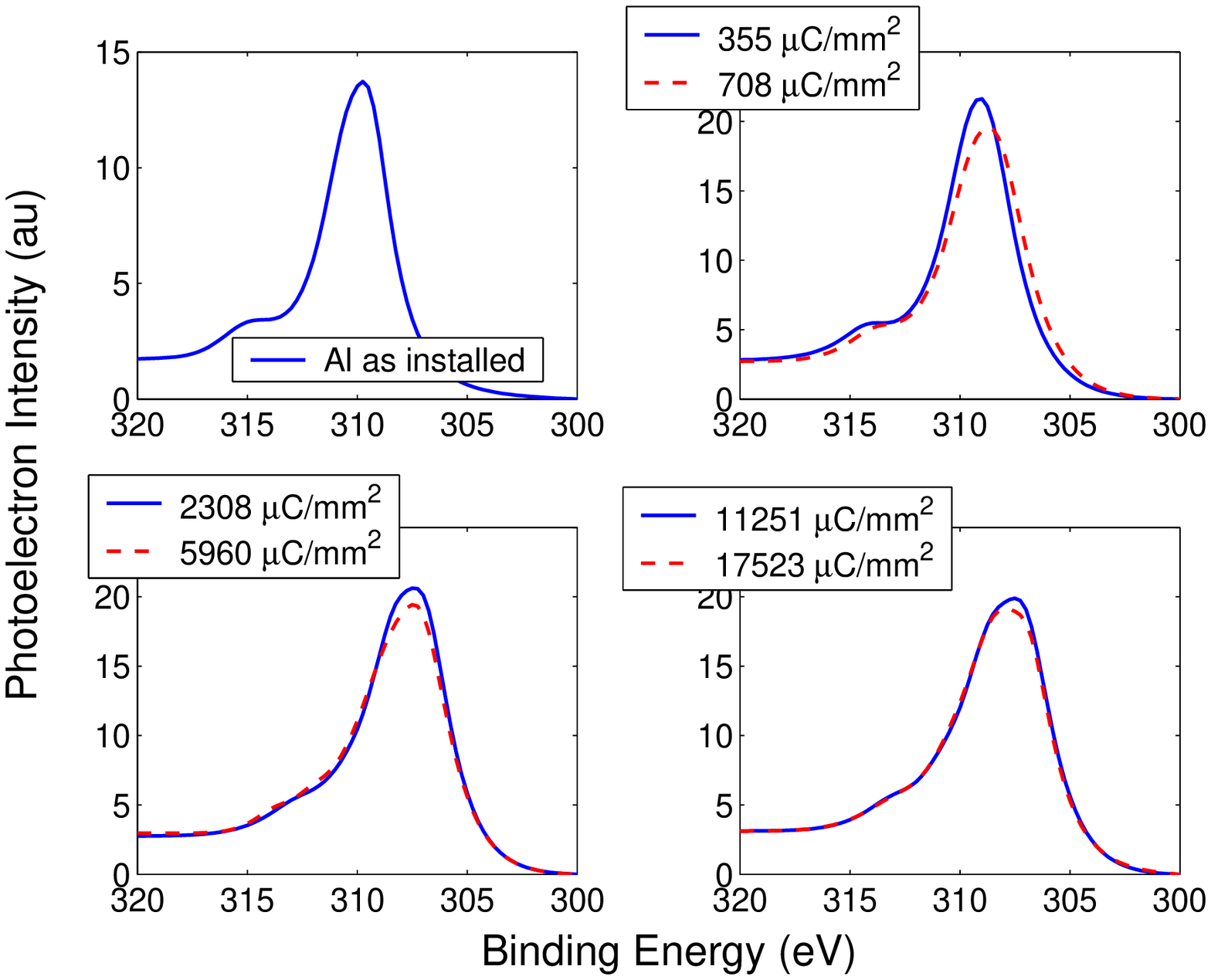}
\caption{Modification of Auger
KL$_{23}$L$_{23}$ Mg peaks (301~eV and 308~eV) during electron
conditioning} \label{figXPSWMgLER6063}
\end{figure}

The Mg KL$_{23}$L$_{23}$ exhibits a shift in energy toward a lower
BE in function of the dose received, Fig.\ref{figXPSWMgLER6063}.
The intensity measured by the electron analyser of the XPS varies
from an as installed piece to the conditioning at a dose of
708~$\mu$C/mm$^2$. From this point on the variation in intensities
are minimal. The Mg2p of this LER sample was also displaced toward
higher BE (54~eV) and shifts toward lower BE (52~eV) during
electron conditioning. Even, if MgCO$_3$ was present on the
surface (BE : 52~eV \cite{XPShandbook:99}) the 54~eV BE cannot be
explained by chemistry alone. In fact, when comparing the binding
energies of several peaks from the LER sample spectrum from the as
installed state versus the 11251~$\mu$C/mm$^2$ electron dosing
state, Fig.\ref{figXPSsurveyLER}, we found that the peaks of the
11251~$\mu$C/mm$^2$ spectrum have shifted 2~eV to lower BE. This
implies that the LER sample as installed must have charge up
during XPS measurement.

\section{Explaining the dip in the SEY curve}

The "as installed" aluminium surface is contaminated by components
in the air, hence a "carbonaceous oxide" layer forms. During
electron conditioning, the ESD process cleans up and modifies the
chemistry of the surface. Unpolymerized hydrocarbons and water are
known to promote a high SEY \cite{Halbritter:84}.
\newline During conditioning the SEY curve goes down,
Fig.\ref{figSEYvsdoseAl}, as modification and removal of the
surface contamination takes place. The SEY curves is the sum of
aluminium-oxide(high SEY) an aluminium surface(low SEY) and
graphitic carbon. At some point, the aluminium surface
contribution prevails, as the oxide layer is not yet formed or
arranged properly. That is the dip of the curves. Past this point,
the contribution of the forming aluminium oxide starts prevailing,
hence raising the SEY.

This model is also supported by others SEY measurements, with a
3~keV electron beam energy, of an evaporated Al and grown
Al$_2$O$_3$ thin film \cite{benka:2002} \cite{benka:2003}. The
SEY, at 3~keV, of an Al thin film exposed to oxygen shows a dip,
and this is independent of the oxygen pressure in the vacuum
chamber \cite{benka:2003}.
\newline For comparison, no dip is seen on copper because copper oxide (Cu$_2$O)
has a lower SEY than the pure Cu \cite{crc} \cite{Bojko:2000}.

Finally, it is not known how much higher doses of electron on the
surface will affect the SEY. It is possible, following our
hypothesis from our XPS observation, that we are building an
aluminium oxide layer, therefore the SEY will keep increasing. The
$\delta_{max}$ could reach any values between 2 and 9, values of
Al$_2$O$_3$ (layer) \cite{crc}.
\newline The usual mechanism for oxidation of metals involves
diffusion of atomic oxygen through the growing oxide layer toward
the underlying metal, which is then oxidized. Thus, the
rate-limiting step for oxide growth, via the Mott-Carbrera
mechanism, is diffusion of oxygen through oxide.

\section{Conclusion}

We have reported on the effects of conditioning, with electrons of
130~eV, on three technical surfaces, aluminium 1100, aluminium
6063 and aluminium 6063 heavily oxidized. We have observed that a
technical aluminium surface does not seem to condition to
saturation with dose, as it is commonly observed for many other
technical surfaces and thin films. The low dose part, below a
mC/mm$^2$, of our results appear normal, Fig.\ref{figSEYvsdoseAl}.
High doses cause oxide growth and the yield rises, contrary to
expected experience. XPS characterization of the chemistry
happening on the surface during conditioning supports our model.

In the framework of the electron cloud problem, the choice of the
technical surface to be used as vacuum chambers is clear.
Non-coated, or otherwise untreated, use of aluminium is a bad
idea, as its conditioned SEY might not go consistently below 2.
However, in an accelerator, ions of few hundred eV can be made
present. Their effect on the surface, from a conditioning
standpoint, is not yet known.

\section{Acknowledgments}
We would like to thanks G. Collet, for his help with the aluminium
nitridation experiment. We also would like to thanks N.~Hilleret
for useful comments.

\clearpage

\clearpage

\clearpage

\clearpage

\clearpage

\clearpage

\clearpage

\clearpage

\clearpage

\clearpage

\clearpage

\clearpage

\clearpage

\clearpage

\clearpage

\clearpage

\clearpage

\clearpage

\clearpage

\clearpage

\clearpage

\clearpage

\clearpage

\begin{figure}[htbp]
\centering
\includegraphics[width=0.7\textwidth,clip=]{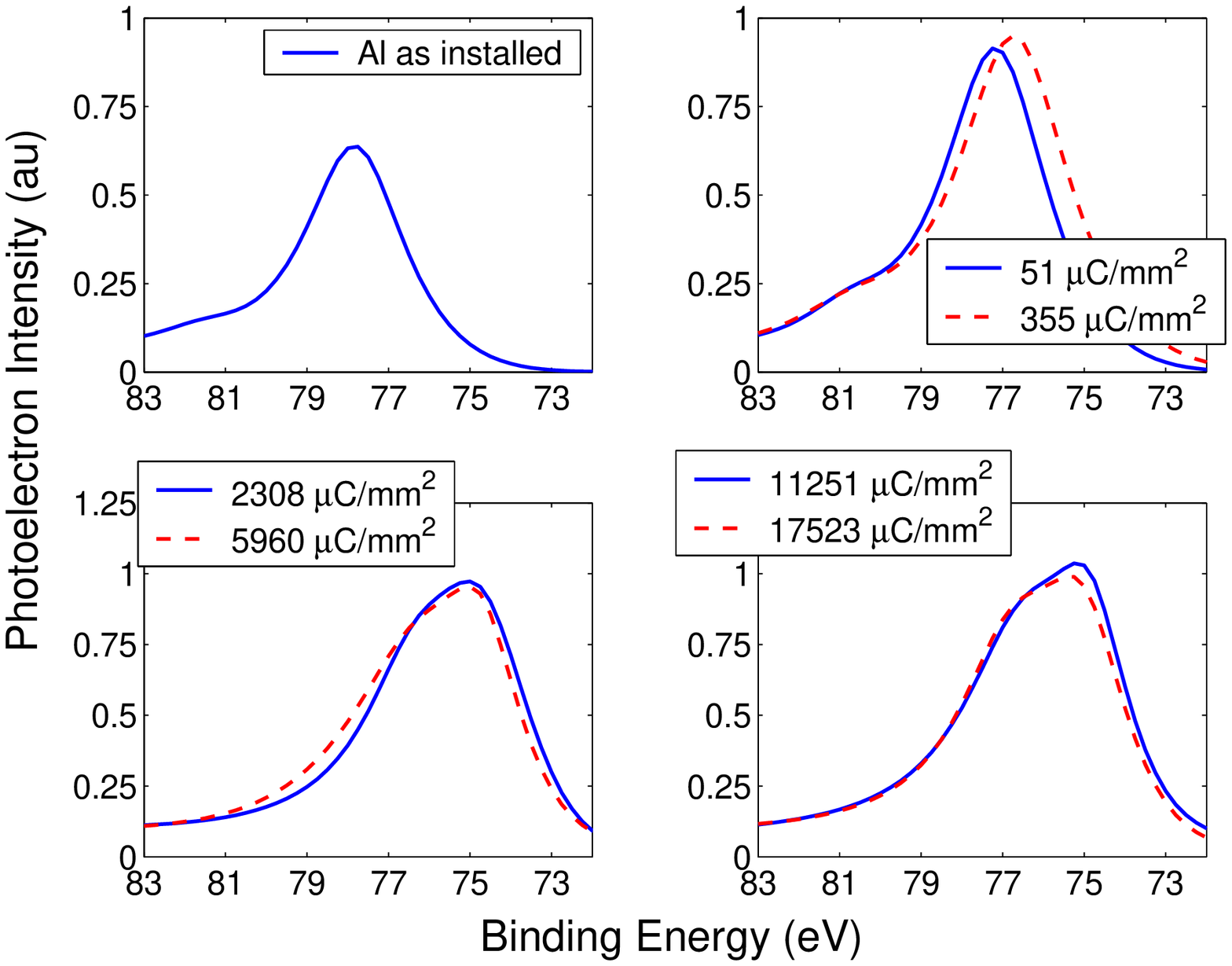}
\caption{Detailed spectra of the Al 2p during electron
conditioning of the LER Al 6063 sample} \label{figXPSWAlLER6063}
\end{figure}

\clearpage

\clearpage

\end{document}